\pdfoutput=1
\documentclass[prc,11pt, amsmath, amsymb, 
aps,reprint,tightenlines,nofootinbib,longbibliography,abbrv,preprintnumbers]{revtex4-1}

\usepackage{graphicx}   
\usepackage[colorlinks=true,linkcolor=blue,citecolor=blue, urlcolor=blue]{hyperref}   
\usepackage{bm} 

\def\st{\begin{equation}}
\def\stp{\end{equation}}
\def\bg{\begin{eqnarray}}
\def\nd{\end{eqnarray}}
\def\Eq#1{Eq.~(\ref{#1})}
\def\Eqs#1{Eqs.~(\ref{#1})}
\def\eq#1{(\ref{#1})}

\def\Fig#1{Fig.~\ref{#1}}

\def\Sect#1{Sect.~\ref{#1}}
\def\Ref#1{Ref.~\cite{#1}}
\def\Refs#1{Refs.~\cite{#1}}

\def\llangle{\left\langle}
\def\rrangle{\right\rangle}

\def\half{\tfrac{1}{2}}

\def\t22{{\scriptscriptstyle T_2T_2}}

\def\xp{{\vec{x}_\perp}}
\def\yp{{\vec{y}_\perp}}
\def\A{{\mathcal A}}
\def\kp{{\vec{k}_\perp}}
\def\chit{{\chi_{\tau_0}^{gg}}}

\def\x{{\bm x}}

\def\k{{\bm k}}
\newcommand{\dlangle}{\langle\!\langle}
\newcommand{\drangle}{\rangle\!\rangle}

\begin{document}

\title{A kinetic regime of hydrodynamic fluctuations and long time tails 
for a Bjorken expansion}

\preprint{INT-PUB-16-018}
\author{Yukinao Akamatsu}
\email{yukinao.akamatsu@stonybrook.edu}
\affiliation{Department of Physics and Astronomy, Stony Brook University, Stony Brook, New York 11794, USA}
\affiliation{Department of Physics, Osaka University, Toyonaka, Osaka 560-0043, Japan}
\author{Aleksas Mazeliauskas}
\email[]{aleksas.mazeliauskas@stonybrook.edu}
\affiliation{Department of Physics and Astronomy, Stony Brook University, Stony Brook, New York 11794, USA}
\author{Derek Teaney}
\email[]{derek.teaney@stonybrook.edu}
\affiliation{Department of Physics and Astronomy, Stony Brook University, Stony Brook, New York 11794, USA}

\date{\today}

\begin{abstract}
We develop a set of kinetic equations for hydrodynamic fluctuations
which are equivalent to nonlinear hydrodynamics with noise.
The hydro-kinetic equations
can be coupled to existing second order hydrodynamic codes to 
incorporate the physics of these fluctuations.
We first show that the kinetic
response precisely reproduces the renormalization of the shear
viscosity and the fractional power ($\propto \omega^{3/2}$) which characterizes
equilibrium correlators of energy and momentum for a static fluid.
Then we use the hydro-kinetic equations to analyze thermal fluctuations
for a Bjorken expansion, evaluating the contribution of 
thermal noise from the earliest moments and at late times. In the Bjorken case, the solution to the  kinetic equations
determines the coefficient of the first fractional power of the gradient 
expansion  ($\propto 1/(\tau T)^{3/2}$) for the expanding system. Numerically, 
we find that the contribution to the longitudinal pressure from hydrodynamic fluctuations  is  
larger than second order hydrodynamics for typical medium parameters 
used to simulate heavy ion collisions. 
\end{abstract}


\maketitle


\section{Introduction}

\subsection{Overview}
The purpose of the current paper is to develop a set of kinetic equations for
hydrodynamic fluctuations, and to use these kinetic equations to study
corrections to Bjorken flow arising from thermal fluctuations.  The specific
test case of Bjorken flow (which is a hydrodynamic model for the longitudinal
expansion of a nucleus-nucleus collision~\cite{Bjorken:1982qr}) is motivated by
the 
experimental 
program of ultra-relativistic heavy-ion collisions at RHIC and the LHC. Detailed  measurements of two particle correlation
functions  have provided overwhelming evidence that the evolution of the excited
nuclear material is remarkably well described by the hydrodynamics of the Quark 
Gluon Plasma (QGP) with a small shear viscosity to entropy ratio of order 
$\eta/s \sim 2/4\pi$~\cite{Heinz:2013th,Luzum:2013yya}.
The typical relaxation times of the plasma, while short enough to support
hydrodynamics, are not vastly smaller than the inverse expansion rates of the collision.  For
this reason
the gradient expansion underlying the hydrodynamic formalism has been
extended to include first and second order viscous corrections~\cite{Baier:2007ix}, and 
these corrections systematically improve the agreement
between hydrodynamic simulations and measured two particle correlations~\cite{Heinz:2013th}.
Additional corrections, which have not been systematically included, arise 
from thermal fluctuations of the local energy and momentum densities and could  be 
significant in  nucleus-nucleus collision where only $\sim 20000$ particles are 
produced. 
This has prompted a keen practical interest in the heavy ion community in
simulating relativistic hydrodynamics with stochastic 
noise~\cite{Gavin:2006xd,Kapusta:2011gt,Yan:2015lfa,Young:2014pka,Murase:2016rhl,Nagai:2016wyx}.
In a non-relativistic context such simulations have reached a fairly mature
state~\cite{bell2007numerical,donev2011diffusive,balboa2012staggered}. 
For a static fluid, thermal fluctuations 
give rise through the nonlinearities of the equations of 
motion to
fractional powers in the fluid response function
at small frequency, $G_{R}(\omega) \propto
\omega^{3/2}$.  Indeed,
the ``long-time tails" 
first observed in molecular-dynamics 
simulations~\cite{velocity_auto,wainwright2,bixon_zwanzig}  are a 
consequence of this non-analytic $\omega^{3/2}$ behavior.
For Bjorken flow, the same nonlinear stochastic physics leads to
fractional powers in the gradient expansion for the longitudinal pressure of
the fluid. One of the goals of this manuscript is to compute the
coefficient of the first fractional power in this expansion.

The measured two particle
correlations in heavy ion collisions reflect both the fluctuations in the initial conditions and
thermal fluctuations. 
Thermal fluctuations are believed to be a small (but conceptually important)
correction to non-fluctuating
hydrodynamics~\cite{Kapusta:2011gt,Young:2014pka,Yan:2015lfa}.  In addition,
thermal fluctuations can become significant close to the QCD critical point~\cite{Stephanov:1998dy} and  in smaller colliding systems such as
proton-nucleus and proton-proton collisions~\cite{Yan:2015lfa}, which show
remarkable signs of collectivity~\cite{Loizides:2016tew}.

In the current manuscript, rather than simulating nonlinear
fluctuating hydrodynamics directly, we will reformulate fluctuating
hydrodynamics as non-fluctuating hydrodynamics (describing a
long wavelength background) coupled to a set of kinetic equations 
describing the phase space distribution of short wavelength hydrodynamic
fluctuations. For Bjorken flow this set of equations can be solved to determine the first fractional powers in the gradient expansion. 

\subsection{Hydrodynamics with noise and fractional powers in the gradient expansion}

Hydrodynamics is a long wavelength effective theory which describes 
the evolution of conserved quantities by organizing corrections 
in powers of gradients.  
For the hydrodynamic expansion to apply we require  
frequencies under consideration to be small compared
to the 
the microscopic 
relaxation rates
\st
\epsilon \equiv  \frac{\omega \eta}{(e+p) c_s^2} \ll 1 
     \label{epsilon}  \, ,
\stp
where we have estimated the microscopic relaxation time with the hydrodynamic 
parameters, $\tau_R\equiv \eta/(e + p) c_s^2$~\cite{Teaney:2009qa} and 
for 
later convenience defined $\epsilon\equiv\omega\tau_R$.

For definiteness, we follow 
precedent~\cite{Son:2007vk,Baier:2007ix,Kovtun:2011np} and consider a conformal 
neutral
fluid driven from equilibrium by a small metric perturbation $h_{xy}(\omega)$ of 
frequency $\omega$. 
Within the framework of linear response (see \Sect{gravitysec} and 
\Ref{Hong:2010at} 
for further details), 
the stress tensor at low frequency 
takes the form 
\st
\label{naiveresponse}
\delta T^{xy} = -h_{xy}(\omega) \Big( p - i\omega \eta 
  + 
  \left(\eta \tau_{\pi} - \frac{\kappa}{2} \right) \omega^2  \Big).
\stp
The first term is the prediction of ideal hydrodynamics $\delta T^{xy} = -p 
h_{xy}$; 
the middle term is the prediction of first order viscous
hydrodynamics~\cite{Son:2007vk}, where $\eta$ is the shear viscosity; 
finally, the last term is the prediction of second order hydrodynamics, where 
$\tau_{\pi}$ and $\kappa$  are 
the associated second order parameters~\cite{Baier:2007ix}.  

In writing 
\Eq{naiveresponse} we have neglected additional contributions stemming from 
fluctuations which will be described below. 
Thermal fluctuations can be incorporated into the hydrodynamic description by 
including
stochastic terms into the equations of 
motion~\cite{LandauStatPart1, LandauStatPart2} 
\begin{align}
d_\mu T^{\mu\nu}=0\label{eomzero},\quad 
T^{\mu\nu}=T^{\mu\nu}_\text{ideal}+T^{\mu\nu}_\text{visc.}+S^{\mu\nu},
\end{align}
where variance of the
noise, $\llangle S^{\mu\nu} S^{\rho\sigma}\rrangle{\sim}2 T\eta \delta(t- t')$,  is determined by the fluctuation dissipation theorem at temperature $T$ and introduces no
new parameters into the effective theory\footnote{
    We follow a standard notation for hydrodynamics summarized in Ref.~\cite*{[see for example: ]Teaney:2009qa}.  $d_{\mu}$ 
    notates a covariant derivative using the ``mostly-plus" metric convention. 
    $T^{\mu\nu}_{\rm ideal} = (e + p) u^{\mu} u^{\nu} + pg^{\mu\nu}$
    and $T^{\mu\nu}_{\rm
    visc} = - \eta \sigma^{\mu\nu}$ where $\sigma^{\mu\nu}  = \Delta^{\mu\rho}
    \Delta^{\nu\sigma} (d_{\rho} u_{\sigma} + d_{\sigma} u_{\rho} - \frac{2}{3}
    g_{\rho\sigma} d_{\gamma} u^{\gamma})$, with $\Delta^{\mu\nu} = g^{\mu\nu} + u^{\mu} u^{\nu}$.
The noise correlator is fully specified in \Eq{noisecorrelator} of \Sect{gravitysec}.}. After including these stochastic
terms, the correlators of momentum and energy evolve 
to their equilibrium values  in the absence of the external force, $h_{xy}(\omega)$. 
Specifically, the
equilibrium two point functions of the energy and momentum densities, 
$\delta e(t,\x)\equiv T^{00}(t,\x) - \llangle T^{00} \rrangle$ and $g^i(t,\x) 
\equiv T^{0i}$ respectively, approach the textbook result~\cite{LandauStatPart1}
\begin{subequations}
\label{etcorrelators}
\begin{align}
 \llangle \delta e(t,\k) \delta e(t,-\k') \rrangle =& \frac{(e + p) T}{c_s^2} 
 \; (2\pi)^3 \delta^3(\k - \k'), \\
 \llangle g^i(t,\k) g^j(t,-\k')) \rrangle =&  (e + p) T \; \delta^{ij}(2\pi)^3 
 \delta^3(\k - \k'),
\end{align}
\end{subequations}
where $c_s$ is the speed of sound, and $\delta e(t,\k)$ notates the spatial 
Fourier transform of $\delta e(t,\x)$.
In the presence of an external force or a non-trivial expansion these correlations are driven 
away from equilibrium. The purpose of hydrodynamics with 
noise is to describe in detail these deviations from equilibrium.

Due to  the nonlinear character of hydrodynamics the thermal
fluctuations change the evolution of the system. Indeed, a diagrammatic
analysis of the hydrodynamic response at one-loop order
shows that the stress in the presence of a weak external field (or the retarded 
Green function) is 
\begin{widetext}
\st
{\llangle T^{xy}(\omega) \rrangle} = -h_{xy}(\omega) \left(p - i \omega 
\eta + (i+1)
\frac{\left(7 + \left(\tfrac{3}{2}\right)^{3/2}\right) }{240\pi}
T \left(\frac{\omega}{\gamma_{\eta}} \right)^{3/2}  + \mathcal
O(\omega^2) \right)  , \label{Txy}
\stp
\end{widetext}
where $p$, $e$, and $\eta$ are renormalized physical quantities 
(see \Sect{grav1} and \Sect{grav2} for further 
discussion of the renormalization),
and
\st
  \gamma_{\eta} \equiv \frac{\eta}{e + p} \, ,
\stp 
is the momentum diffusion coefficient
~\cite{Kovtun:2003vj,Kovtun:2011np}.
As emphasized and estimated previously,
the fractional order $\omega^{3/2}$ is parametrically  larger 
than second order hydrodynamics~\cite{Kovtun:2011np}. However, the coefficient of 
the $\omega^{3/2}$ terms is vanishingly small in weakly coupled theories
and in strongly coupled theories at large $N_c$, and 
therefore second order hydrodynamics may be an effective approximation scheme  
except at very small frequencies.
In the context of holography,
the $\omega^{3/2}$ term can only be determined by performing a one 
loop calculation in the bulk~\cite{CaronHuot:2009iq}. 

In the current paper we will rederive \Eq{Txy}  using a kinetic description of 
short wavelength hydrodynamic fluctuations.
For an external driving frequency of order
$\omega$,
we identify an important length scale set by equating  the damping rate and  
the external frequency
\st
\label{kstarw}
\gamma_{\eta}  k_*^2 \sim \omega \, ,\quad   k_{*} \sim  
      \left(\frac{ 
      \omega  }{\gamma_\eta} \right)^{1/2} \, .
\stp
We will refer to the $k_*$ as the \emph{dissipative scale} below (see also
\Ref{CaronHuot:2009iq}).
Modes with wavenumbers significantly larger than the dissipative scale, $k\gg k_{*}$,  are  damped
and reexcited by the noise on a time scale which is short compared to period 
$2\pi/\omega$, and
this rapid competition leads to the equilibration of these shorter wavelengths, i.e. 
their equal  time correlation functions are given by \Eq{etcorrelators}. 
By contrast, modes with wavenumbers of order $k\sim k_{*}$ have equal time 
correlation functions which deviate from the equilibrium expectation values.

\begin{figure}
\includegraphics[width=1\linewidth]{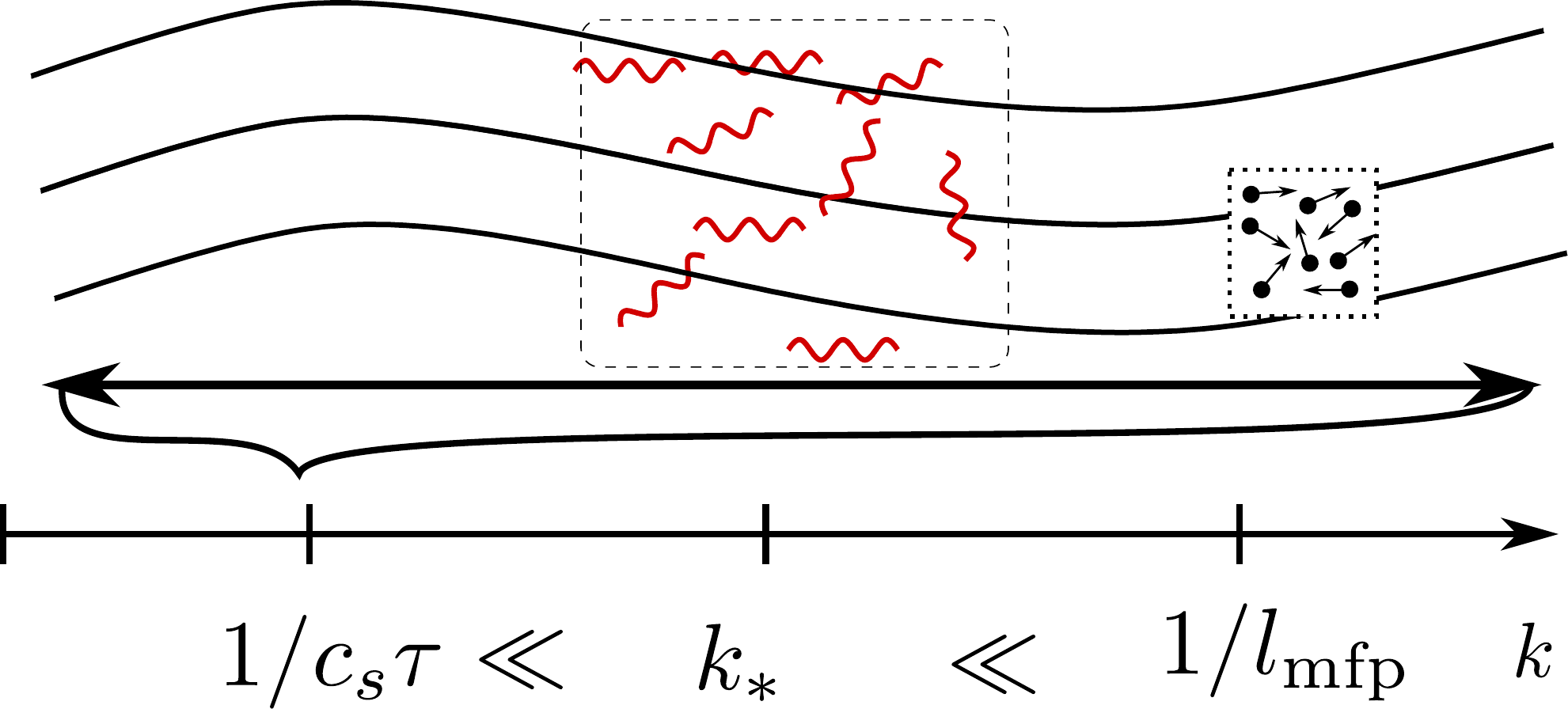}
\caption{\label{scales} The hydro-kinetic description of noise is 
based on 
the  separation of scales between the long wavelength hydrodynamic background 
(with $k \sim 1/c_s\tau$), and shorter wavelength hydrodynamic
fluctuations (with $k \sim k_{*}\equiv 1/\sqrt{\gamma_\eta\tau}$). 
The wavelengths of the hydrodynamic
fluctuations  are still much longer than microscopic mean free path. 
The hydrodynamic fluctuations are driven out of
equilibrium by the expanding background, and this 
deviation is the origin of the long-time tail correction to the stress tensor. }
\end{figure}
It is notable that the wavenumbers of interest $k_*$ are large 
compared to $\omega/c_s$, 
but still small compared to microscopic
wavenumbers of order the inverse mean free path.  Estimating the
mean free path as $\ell_{\rm mfp} = c_s\tau_R$, we see that
the strong inequalities
\st
\label{scalesep0}
\frac{\omega}{c_s} \ll k_{*}  \ll  \frac{1}{\ell_{\rm mfp} },
\stp
can be written as 
\st
\label{scalesep}
\frac{\omega}{c_s} \ll  \frac{\omega}{c_s} \, \frac{1}{\sqrt{\epsilon}} \ll \frac{\omega}{c_s} \, \frac{1}{\epsilon} ,
\stp
and thus holds whenever hydrodynamics is applicable, $\epsilon \ll 1$.
The scale separation 
illustrated in \Fig{scales}
can be used to set up an 
approximation scheme where modes of order $k_*$ are treated with
a kinetic or WKB type approximation scheme. We will develop the appropriate 
kinetic equations in \Sect{gravitysec}.  These kinetic equations
can be solved and used to determine how the two point functions of 
energy and momentum with wavenumbers of order $k_*$ deviate from equilibrium
when driven by an external perturbation. The $\omega^{3/2}$ term
in \Eq{Txy} roughly  represents  the contribution of $\int k^2 dk  {\sim}k_*^3$ 
slightly out of equilibrium hydro-kinetic modes per volume, with each mode 
contributing  $\half T$ of energy to the stress tensor. Note that the 
contribution to the 
stress tensor of 
modes outside of the kinetic regime $k\ll k_*$  is suppressed by phase 
space.

Similar kinetic equations can be derived for much more general flows. 
We will establish the appropriate kinetic equations for a Bjorken
expansion~\cite{Bjorken:1982qr}, which is a useful model for the 
early stages of a heavy ion collision. 
The ideal, first, and second order terms in the gradient expansions
have been given in Refs.~\cite{Bjorken:1982qr},\; \cite{Danielewicz:1984ww},\; and \cite{Baier:2007ix,Bhattacharyya:2008jc} respectively. 
For a conformal (non-fluctuating) fluid the longitudinal pressure during a 
Bjorken expansion takes the form
\st
\tau^2T^{\eta\eta} = p - \frac{4}{3} \frac{\eta}{\tau} +  \frac{8}{9\tau^2} (\lambda_1 - \eta\tau_\pi) + \ldots  \, .
\stp
The expansion rate is $\partial_{\mu} u^{\mu} = 1/\tau$, and each
higher term in the gradient expansion is suppressed by an 
integer power of $1/\tau T$.
For  Bjorken flow the expansion rate plays the role of frequency, and
the distribution of sound modes are characterized  by a dissipative scale analogous
to \Eq{kstarw} of order\footnote{
    The quantities $k_{*}(\tau)$, $\gamma_{\eta}(\tau)$, $s(\tau)$, $\ldots$   are
    all functions of time for a Bjorken expansion, e.g.  for a conformal equation of state and an ideal expansion, $k_{*}(\tau)
    \propto \tau^{-2/3}$, $\gamma_\eta \propto \tau^{1/3} $, $s(\tau) \propto
    \tau^{-1}$,  etc.  Throughout the paper $k_{*}$,$\gamma_{\eta}$, $s$,
    $\ldots$ (without a time argument) will denote the physical quantity
    at the final time of consideration. The explicit time argument will 
    be used when needed, e.g. $k_{*}(\tau') = k_{*} (\tau/\tau')^{-2/3} $.
}
\st
k_{*}   \sim \frac{1}{(\gamma_\eta \tau)^{1/2}} 
.\label{kstar}
\stp
At this scale the viscous damping rate balances the expansion rate. These hydrodynamic modes satisfy the inequality
\st
\label{scalesep2}
\frac{1}{c_s \tau} \ll  k_{*}  \ll \frac{1}{\ell_{\rm mfp} }, 
\stp
and this strong set of inequalities can be used to determine 
a kinetic equation for hydrodynamic modes of order $k_*$. The equal
time correlation functions for wavenumbers of this order deviate 
from their equilibrium form in \Eq{etcorrelators}, and the kinetic equations precisely determine
the functional dependence of this deviation. Finally, these
modes contribute to the longitudinal pressure and determine
first fractional power  in the longitudinal pressure of a conformal fluid (analogous to \Eq{Txy}). In \Sect{bjorken} we will establish that 
this nonlinear correction to the longitudinal component of the 
stress tensor is
\begin{multline}
\frac{\llangle \tau^2T^{\eta\eta}\rrangle}{e + p} =   \Big[ \frac{p}{e+p} - \frac{4}{3} 
 \frac{\gamma_\eta}{\tau}
+   \frac{1.08318}{s\, (4\pi \gamma_\eta \tau)^{3/2} } + \\  \mathcal O\left( \frac{1}{(\tau T)^2} \right) \Big].
\end{multline}
Noise also contributes to transverse momentum 
fluctuations, and this contributes at quadratic order to $\llangle T^{\tau\tau} \rrangle$ as we discuss
in \Sect{bjorken}. Thus,
a complete description of a Bjorken expansion with noise must also reexamine the 
relationship between the background energy density $e$, and the one point function $\llangle T^{\tau\tau} \rrangle$.

An outline of the paper is as follows. In \Sect{gravitysec} we consider a static fluid perturbed by an 
external gravitational perturbation. The purpose of this section is to introduce
the kinetic equations, and to reproduce
the results of the diagrammatic analysis of Refs.~\cite{Kovtun:2003vj,Kovtun:2011np} using the hydro-kinetic
theory adopted here.  In \Sect{bjkinetic} we linearize the hydrodynamic equations
of motion to determine the appropriate kinetic equations for a Bjorken 
expansion. In \Sect{bjstress} and \ref{sketch} we determine the solutions to 
the kinetic theory and use these solutions to evaluate the contribution of 
hydrodynamic modes to the stress tensor. We give an intuitive physical 
interpretation of the main results of the paper in 
\Sect{qualitative}. 
Finally we 
conclude with results and discussion in \Sect{results}.


\section{Hydrodynamic fluctuations in a static fluid}
\label{gravitysec}

We will first derive the kinetic equations for 
hydrodynamic fluctuations in homogeneous flat space in \Sect{grav1}.
The purpose here is to introduce notation, and  to discuss the 
kinetic approximations in the simplest context. Then in \Sect{grav2}
we will perturb the system with a gravitational field and derive 
the appropriate kinetic theory in this case. 
We then use this hydro-kinetic theory to reproduce the results of 
loop calculations~\cite{Kovtun:2003vj,Kovtun:2011np}
for the renormalization of the shear viscosity 
and the long-time tails which characterize the hydrodynamic response due to 
nonlinear noise effects.

\subsection{Relaxation equations for hydrodynamic fluctuations}
\label{grav1}

To illustrate the approximations that follow and to introduce
notation, 
we first will derive 
kinetic equations for the two point functions for energy 
and
momentum density perturbations around a static homogeneous background. 
The (bare) background quantities of the hydrodynamic effective theory, such as the energy density, pressure, and shear
viscosity ($e_0(\Lambda)$, $p_0(\Lambda)$, and $\eta_0(\Lambda)$ respectively) are calculated
by integrating out fluctuations above a scale $\Lambda$, i.e. by excluding the contributions of hydrodynamic fluctuations with wavenumber $k < \Lambda$ to the stress tensor.  This is important because modes with $k < \Lambda$ will not 
be in equilibrium when the system is perturbed by a driving force.
The relation between the bare parameters and the physical quantities (which may be computed in infinite volume with lattice QCD for instance) is discussed in \Sect{grav2} and in \Ref{Kovtun:2011np}, where $\eta_0(\Lambda)$ is referred to as $\eta_{\rm cl}(p_{\rm max})$.

To derive 
a relaxation equation for the two point functions we linearize the equations of stochastic hydrodynamics and
study the eigenmodes of the system. The correlations between eigenmodes
with vastly different frequencies are neglected in a kinetic
(or coarse graining) approximation.
For the constant background $e_0=\text{const}$, and to linear order in 
field perturbations and stochastic
fluctuations, the equations of motion (\Eq{eomzero}) become
\begin{subequations}
\label{eomflat}
\begin{align}
\partial_t \delta e +ik_i g^i &=0,\label{eom21}\\
 \partial_t g_{i} +i k_i \delta p+\gamma_\eta k^2 
g_i+\frac{1}{3}\gamma_\eta k_ik_j g^j
&=-\xi_i\label{eom22},
\end{align}
\end{subequations}
where $\gamma_\eta \equiv \eta_0/(e_0+p_0)$ is
computed with bare quantities, 
and ${-}\xi_i$ is the stochastic force, ${-}ik_j 
S^{j}_{\phantom{k}i}(t,\k)$. Here $S^{j}_{\phantom{k}i}(t,\k)$ are spatial 
components of the noise tensor 
with equilibrium correlation given by~\cite{LandauStatPart2}
\begin{widetext}
\begin{align}
    \label{noisecorrelator}
&\langle S^{\mu\nu}(t_1,\k)S^{\alpha\beta}(t_2,-\k')\rangle =
2T \eta_0\left[
\left(\Delta^{\mu\alpha}\Delta^{\nu\beta} + 
\Delta^{\mu\beta}\Delta^{\nu\alpha}\right)
- \frac{2}{3}\Delta^{\mu\nu}\Delta^{\alpha\beta}
\right](2\pi)^3\delta^3(\k-\k') \delta(t_1 - t_2).
\end{align}
\end{widetext}
 It is convenient to combine 
\Eq{eomflat}  into a single matrix 
equation for an amalgamated field $\phi_a=(c_s \delta e, g_j)$
\begin{equation}
\partial_t \phi_{a}(t, \k)=-i \mathcal{L}_{ab} \phi_b-\mathcal{D}_{ab} \phi_b 
- \xi_a 
\label{eom3},
\end{equation}
where ideal and dissipative terms are 
\begin{equation}
 \mathcal{L}_{ab}=\begin{pmatrix}0 &c_s k_j \\c_sk_i & 0 \end{pmatrix},\quad  
 \mathcal{D}_{ab}=\gamma_\eta\begin{pmatrix}0 &0\\0 & k^2 
 \delta_{i  j}+\frac{1}{3}k_i k_j
 \end{pmatrix},
\end{equation}
and the stochastic noise $\xi_a$ satisfies correlation equation
\begin{multline}
    \left<\xi_a(t_1,\k) \xi_b(t_2,-\k')\right>= 2T 
    (e_0+p_0)\mathcal{D}_{ab}  (2\pi)^3  \\ \times  \delta^3(\k-\k')   \delta(t_1-t_2).\label{xixi}
\end{multline}
At the dissipative scale
the acoustic matrix $\mathcal{L}\sim c_sk_*$ 
originating from ideal equations of motion dominates over the competing 
dissipation  $\mathcal{D}$ and 
fluctuation $\xi_a$ terms. $\mathcal 
L_{ab}$   has 
four 
eigenmodes: two 
longitudinal 
sound modes 
with $\lambda_{\pm} = \pm c_s |\k|$  and two transverse 
zero modes ($\lambda_{T_1}= \lambda_{T_2} = 0$). 
Since $\mathcal{L}$ drives evolution of $\phi_a$, it will be convenient to 
analyze the dynamics in terms of eigenmodes 
of $\mathcal L_{ab}$:
\begin{equation}
   (e_{\pm})_a = \frac{1}{\sqrt{2}}
\begin{pmatrix}
    1 \\
  \pm\hat{k}
\end{pmatrix},
\quad
(e_{T_1})_a =
\begin{pmatrix}
    0 \\
    \vec{T}_1 \\
\end{pmatrix},
\quad
(e_{T_2})_a =
\begin{pmatrix}
    0 \\
    \vec{T}_2 \\
\end{pmatrix},\label{vectors}
\end{equation}
where $\hat{\k}=\k/|\k|$, and $\vec{T}_1$  and  $\vec{T}_2$ are
two orthonormal spatial vectors perpendicular to $\hat\k$
\begin{subequations}
\begin{align}
\hat{k}&=(\sin \theta \cos \varphi, \sin \theta \sin \varphi, \cos \theta),\\
\vec{T}_1&=(-\sin \varphi, \cos \varphi, 0),\\
\vec{T}_2&=(\cos \theta \cos \varphi, \cos \theta \sin \varphi, -\sin \theta).
\end{align}
\end{subequations}
Now we will derive a relaxation equation for the two point correlation 
function of hydrodynamic fluctuations by defining a density matrix 
$N_{ab}(t,\k)$
\begin{equation}
\left<\phi_a(t,\k) 
\phi_b(t,-\k')\right>\equiv 
N_{ab}(t,\k)(2\pi)^3\delta^3(\k-\k'),\label{defN}
\end{equation}
and analyzing the time evolution of $N_{ab}(t,\k)$.

The analysis is most transparent in the eigenbasis, $\phi_{A} \equiv \phi_a \left(e_{A}\right)_a$ with $A=+,-,T_1, T_2$,
and below we will determine the equation of motion for  $N_{AB}\equiv\llangle \phi_A \phi_B\rrangle $ where $A,B=+,-,T_1,T_2$.
We note that the positive and negative sound modes $\phi_{+}$ and $\phi_{-}$
are related since the hydrodynamic
fields are real, $\phi_{-}^*(\k,t) = \phi_{+}(-\k,t)$.

Using the equations of motion for $\phi_A$ we calculate the infinitesimal change of 
$N_{AB}(t+\Delta t)-N_{AB}(t)$,  and use the equal time correlator for the 
noise  (\Eq{xixi}) to find a differential equation for $N_{AB}$ 
\begin{equation}
\partial_t N=-i[\mathcal{L},N] 
-\{\mathcal{D}, N\}+2T (e_0+p_0)\mathcal{D},\label{eomAB}
\end{equation}
where $[X,Y]\equiv XY-YX$, $\{X,Y\}\equiv XY+YX$, and $[\mathcal{L},N]_{AB} =(\lambda_A-\lambda_B) 
N_{AB}$.
We are interested in the evolution of two point 
correlation functions over time scales much larger than acoustic oscillations,
$\Delta t \gg 1/(c_s k_*)$. 
On these timescales the off-diagonal matrix elements
of the density matrix, $N_{+T_1}$ for example,  
rapidly oscillate reflecting the large difference in eigenvalues, $\lambda_{+} 
- \lambda_{T_1} \sim c_s k_*$.
In a coarse graining approximation the contributions of 
these off-diagonal matrix elements to physical quantities can be neglected when 
averaged over times
long compared to $1/(c_s k_*)$.  
This reasoning does 
not apply to the diffusive modes 
$A,B=T_1,T_2$ where both eigenvalues are zero,  but rotational symmetry in the 
transverse $xy$-plane requires $N_{T_1T_2}$ to vanish\footnote{Rotational 
symmetry in the 
transverse $xy$-plane requires that $\left<g_i g_j\right>\sim A \delta_{ij} + B 
\hat{k}_i \hat{k}_j$, where $i,j=x,y$. Such a tensor structure has vanishing 
$T_1T_2$ projection.}.

With these approximations, the non-trivial relaxation equations of two point 
correlation functions in \Eq{eomAB} are
\begin{subequations}
\label{Nflat}
\begin{align}
\partial_t N_{\pm\pm}(t,\k)  &= -\frac{4}{3}\gamma_\eta k^2 
    (N_{\pm\pm}-N_0)\label{N++},\\
\partial_t N_{T_1T_1}(t,\k)  &= -2\gamma_\eta k^2 
(N_{T_1T_1}-N_0),\\
\partial_t N_{T_2T_2}(t,\k)  &= -2\gamma_\eta k^2 
(N_{T_2T_2}-N_0)\label{NT2T2},
\end{align}
\end{subequations}
where 
\st
N_0=T (e_0+p_0)
\stp 
is the equilibrium value for $N_{AA}$ 
(c.f. \Eq{etcorrelators}). In the absence of external perturbations, two point 
correlation functions relaxes to their equilibrium values.
The next step towards general kinetic equations is to study how equal time 
correlations are driven out of equilibrium by the presence of external fields.

\subsection{Linear response to gravitational perturbations}
\label{grav2}
In this section we will study the evolution of two point energy and momentum correlators in the presence of time varying gravitational field. We determine
the kinetic equations in the time dependent background, and use these
equations to reproduce the modifications of
the retarded Green function (\Eq{Txy}) due to thermal 
fluctuations, which were previously found by a one-loop calculation~\cite{Kovtun:2003vj,Kovtun:2011np}.

A straightforward way of introducing an external source to equations of motion 
is to study fluctuating hydrodynamics in the presence of a small metric 
perturbation, $g_{\mu\nu} = \eta_{\mu\nu} + h_{\mu\nu}$.
The Green function records the response of $T^{\mu\nu}$ to the metric perturbation
\begin{equation}
\delta \llangle T^{\mu\nu}(\omega) \rrangle = 
-\frac{1}{2}G^{\mu\nu,\alpha\beta}_R(\omega)h_{\alpha\beta}(\omega).
\end{equation}

For a constant homogeneous background with time dependent metric perturbation $h_{ij}(t)$, symmetry 
constrains the form of the retarded Green function
\begin{multline}
    G_R^{ij, kl}(\omega)= 
    \mathring{G}_R(\omega)\, ( \delta^{ik}\delta^{jl}+ \delta^{il}\delta^{jk}-\frac{2}{3}\delta^{ij}\delta^{kl}) + \\
\overline{G}_{R}(\omega)\, \delta^{ij}\delta^{kl}\label{Gsym},
\end{multline}
and therefore we can obtain the Green function in \Eq{Txy}, i.e. 
$\mathring{G}_R(\omega)$, 
by considering a 
diagonal traceless metric perturbation,
$h_{ij}(t)=h(t)\, \text{diag}\,(1,1,-2)$.

In the presence of metric perturbations and thermal fluctuations, the energy 
momentum tensor is 
\begin{equation}
    \label{tijeq}
    \delta \llangle T^{ij}(t)\rrangle = -p_0h^{ij}-\eta_0  \partial_th^{ij} 
+ \frac{\left<g^i(t,\x) g^j(t,\x)\right>}{e_0+p_0},
\end{equation}
where the nonlinear term stems from the constitutive relation of 
ideal hydrodynamics, $T^{ij} = p_0 \delta^{ij} + (e_0 + p_0) u^{i} u^{j}$.
The averaged squared momentum, $\llangle g^i(t,\x) g^j(t,\x)\rrangle$, is related to the two-point 
functions of $g^{i}$ in $\k$ space as
\begin{equation}
\left<g^i(t,\x) g^j(t,\x)\right> =\int 
\frac{d^3k}{(2\pi)^3}N^{ij}(t,\k).\label{gg}
\end{equation}
In this integral, the equilibrium value of $N^{ij}$ and its first viscous correction will 
renormalize $p_0$ 
and $\eta_0$ (see below), while the finite remainder 
will determine the first fractional power in the stress tensor correlator $\propto \omega^{3/2}$.

Studying the hydrodynamic equations in \Eq{eomzero}, and neglecting  metric perturbations of the dissipative terms, 
we find  that the
linearized equations of motion  are identical to flat background \Eq{eomflat}, but now 
there is a difference between covariant and contravariant indices
\begin{subequations}
\begin{align}
\partial_t \delta e +ik_i g^i &=0,\\
 \partial_t g_{i} +i k_i \delta p+\gamma_\eta k^2 
g_i+\frac{1}{3}\gamma_\eta k_ik_j g^j
&=-\xi_i.
\end{align}
\end{subequations}
To avoid this complication,
we use a  vielbein formalism and scale  the spatial components of momentum and wavenumber by 
$\sqrt{g_{ij}}$, i.e.  
$g^i$ and $k_j$ are replaced by
\begin{subequations}
\begin{align}
G_{\hat{\imath}}&=(1+\frac{1}{2}h_{ij})g^j,\\
K_{\hat{\imath}}&=(1-\frac{1}{2}h^{ij})k_j,
\end{align}
\end{subequations}
where now the position of hatted indices is unimportant.
Analogously 
to \Eq{eom3}, we obtain a matrix equation for $\phi_a=(c_s \delta e, G_{\hat 
\imath})$
\begin{equation}
\partial_t \phi_{a}(t, \k)=-i \mathcal{L}_{ab} \phi_b-\mathcal{D}_{ab} \phi_b - 
\xi_a -
\mathcal{P}_{ab}\phi_b,
\label{eom4}
\end{equation}
with an additional metric dependent source term   
\begin{equation}
 \mathcal{P}_{ab}=\begin{pmatrix}0 &0 \\0 & \frac{1}{2}\partial_0 h_{\hat{i} 
 \hat{j}} 
 \end{pmatrix},
\end{equation}
which drives the hydrodynamic fluctuations away from equilibrium.
The eigenbasis of $\mathcal{L}$ (see \Eq{vectors}) is now defined with respect 
to the time dependent vector
$\vec{K}(t)$, but remains orthonormal at all times. Furthermore, the metric 
perturbation preserves rotational symmetry in the transverse $xy$-plane, and 
this 
guarantees that the $T_1$ and $T_2$ modes are not mixed by the time-dependent perturbation. Thus, the only non-trivial 
diagonal components of the symmetrized energy and momentum two point functions  are 
\begin{subequations}
    \begin{multline}
\partial_t N_{\pm\pm}  = -\frac{4}{3}\gamma_\eta K^2 
(N_{\pm\pm}-N_0)  \\
  -\frac{1}{2}\partial_t h \, (\sin^2\theta_K-2\cos^2 \theta_K)N_{\pm\pm}, 
  \end{multline}
  \begin{align}
\partial_t N_{T_1T_1}  &= -2\gamma_\eta K^2 (N_{T_1T_1}-N_0)-\partial_t h \,
N_{T_1T_1},
\end{align}
\begin{multline}
\partial_t N_{T_2T_2}  = -2\gamma_\eta K^2 (N_{T_2T_2}-N_0) \\
   -\partial_t h \,
(\cos^2 \theta_K-2\sin^2\theta_K) N_{T_2T_2}.
\end{multline}
\end{subequations}
We can find a perturbative solution to these equations for a small periodic metric perturbation, e.g.
\begin{equation}
    N_{T_2T_2}(\omega,\k) \simeq N_0\left(2\pi\delta(\omega){+}\frac{i\omega 
    h(\omega) (\cos^2\theta_K {-} 2 \sin^2\theta_K)}{-i\omega + 2\gamma_\eta 
    K^2}\right).
\end{equation}
To find the correction to the energy momentum tensor due to the nonlinear momentum 
fluctuations in \Eq{tijeq}, we need to perform the $k$ 
space integral in \Eq{gg}
\begin{align}
\langle\phi_a(x)\phi_b(x)\rangle
=&{\int} \frac{d^3 K}{(2\pi)^3} N_{ab}(\tau,\k),\nonumber \\
=& {\int} \frac{K^2 dK d\cos \theta_K d\varphi_K}{(2\pi)^3}  \nonumber \\
  &   \qquad \times (e_A)_a N_{AB}(\tau,\k) (e_B)_{b} \, .
\end{align}
Note, care should be taken when transforming the zeroth 
order value $N_{AA}=N_0$ to original unhatted basis as it produces terms linear 
in 
metric perturbation.
The 
modification 
of the response function $\mathring{G}_R(\omega)$ due to the momentum fluctuations (i.e. the last term in \Eq{tijeq}) is
\begin{align}
    \mathring{G}_R(\omega) &= -\frac{1}{6}(\delta T^{xx}+\delta T^{yy}-2\delta 
T^{zz})/h(\omega),\nonumber\\
&\supset -\frac{T}{6}\int\frac{d^3 K}{(2\pi)^3}
\left(-6+  i\omega  
 \frac{(\sin^2\theta_K-2\cos^2\theta_K)^2}{-i\omega +\frac{4}{3}\gamma_\eta 
 K^2}\right. \nonumber \\
& \quad\qquad \left.  
 + \; \;  
 i\omega\frac{1+(\cos^2\theta_K-2\sin^2\theta_K)^2}{-i\omega+2\gamma_\eta 
 K^2}\right).
\end{align}
Performing $K$-space integral with UV cutoff, $K_\text{max}=\Lambda$, and
adding the remaining terms in \Eq{tijeq},
we find
\begin{multline}
\label{flatans}
    \mathring{G}_R(\omega)  =
    \left( p_0 + \frac{\Lambda^3}{6\pi^2} T \right)
        {-}i \left(\eta_0  + \frac{\Lambda}{\gamma_\eta} \frac{17}{120\pi^2}T 
        \right) \omega  \\
 +(1+i)\frac{1}{\gamma_\eta^{3/2}}  
\frac{ (\frac{3}{2})^{3/2}
+ 7 }{240\pi} T\omega^{3/2},
\end{multline}
in agreement with previous work~\cite{Kovtun:2003vj,Kovtun:2011np}.
The first two terms in \Eq{flatans} are the 
renormalized pressure ($p \equiv p_0(\Lambda) + O(T\Lambda^3)$) and shear
viscosity ($\eta \equiv \eta_0(\Lambda) + O(\Lambda T^2)$) as discussed 
previously~\cite{Kovtun:2011np}. Further
discussion of the renormalization of these quantities is given in the next section when the expanding
case is presented.

The last term is the finite nonlinear modification of the medium response,
and agrees with loop calculations in equilibrium. The
kinetic approach outlined in this section has the advantage that it can be readily applied to more general backgrounds, and we will exploit this advantage to calculate the analogous correction for a Bjorken 
expansion in the next section.  In contrast to the linear response described here,
the deviation from equilibrium in the expanding case is of order unity. 
Consequently, computing the first fractional power in an expanding system
with the diagrammatic formalism would require
an extensive resummation, which would invariably reproduce kinetic calculation 
described 
in the next section~\cite{Jeon:1995zm}.


\section{Hydrodynamic fluctuations for a Bjorken expansion}
\label{bjorken}

In this section we will derive the kinetic evolution equations for hydrodynamic fluctuations during 
a 
Bjorken expansion.
We consider a neutral conformal fluid, for which $c_s^2=1/3$, $\zeta=0$, and $\mu_{\rm B}=0$.
In Bjorken coordinates
the energy and momentum conservation laws are
\bg
\partial_{\mu}T^{\mu\nu}+\frac{1}{\tau}T^{\tau\nu} + \Gamma^{\nu}_{\mu\beta}T^{\mu\beta}=0,
\nd
with $\Gamma^{\tau}_{\eta\eta}=\tau$ and $\Gamma^{\eta}_{\tau\eta}=\Gamma^{\eta}_{\eta\tau}=1/\tau$~\cite*{[see for example: ]Teaney:2009qa}.
For hydrodynamics without noise  the background
flow fields  are independent of transverse coordinates
and rapidity and satisfy 
\begin{align}
    \frac{d (\tau T^{\tau\tau})}{d\tau} &= - \tau^2T^{\eta\eta}, \\
    \frac{d (\tau T^{\tau i})}{d\tau} &= 0,
\end{align}
where roman indices, $i,j \ldots $, run over transverse coordinates $x,y$.
The transverse momentum $T^{\tau i}$ is constant, and can be
chosen to be zero. In hydrodynamics
$T^{\tau\tau}$ and $\tau^2 T^{\eta\eta}$  are related by 
constitutive equations 
\begin{align}
    T^{\tau\tau} &= e ,\\
    \tau^2T^{\eta\eta} &= c_s^2 e - \frac{4\eta}{3\tau}.
\end{align}
Note  that in $\tau^2 T^{\eta\eta}$ the viscous correction is of order 
$\epsilon=\eta/(e+p)\tau\ll1$ smaller than the ideal part, and the solution is 
approximately $e(\tau) = e(\tau_0)\cdot (\tau_0/\tau)^{1+c_s^2}[1+\mathcal O(\epsilon)]$.  

We will consider 
the evolution of linearized fluctuations on top of this background.  
The effect of these fluctuations on the background evolution
can then be included as a correction after the two point functions 
are known, i.e.
\st
\frac{d \dlangle T^{\tau\tau} \drangle}{d\tau} = -\frac{ \dlangle T^{\tau\tau}\drangle  +  \dlangle \tau^2T^{\eta\eta} \drangle}{\tau},\ \ \
\stp
where  the constitutive  relations take the form
\begin{align}
    \label{eave}
    \dlangle T^{\tau\tau} \drangle = e  +  \frac{\dlangle \vec{G}^2 \drangle}{e +p} , \\
    \label{pLave}
     \dlangle \tau^2T^{\eta\eta} \drangle = c_s^2 e - \frac{4\eta}{3\tau}  + 
     \frac{\langle\!\langle (G^{\hat z})^2 \rangle\!\rangle}{e +p}.
\end{align}
Here and below $e(\tau)$ is the average rest frame energy density\footnote{
    $e(\tau)$ notates the average \emph{rest frame} energy density 
    and does not fluctuate; $\dlangle T^{\tau\tau}\drangle$ is
    the average energy density.
In general, the rest frame energy density $e + \delta e$ in a finite volume 
would be estimated from sample estimate of $T^{\tau\tau}$ and
$\vec{G}$  through the (ideal) constitutive
equations, $e + \delta e \simeq T^{\tau\tau} -  \frac{\vec{G}^2}{(1 + c_s^2) 
T^{\tau\tau}}$. Thus $e$ is given by \Eq{eave}, and $\delta e \simeq \delta 
T^{\tau\tau} - \delta (\vec{G^2}/T^{\tau\tau})/(1+c_s^2) \simeq \delta 
T^{\tau\tau}$.};
$\vec{G}$ is
the momentum density $\vec{G} = (T^{\tau x}, T^{\tau y}, \tau T^{\tau \eta})$, 
and all quantities are renormalized as explained more completely below.

There are two sorts of fluctuations to consider: fluctuations in 
the initial conditions (which are long range in rapidity), and hydrodynamic
fluctuations stemming from thermal noise (which are short range in rapidity). The
average over the initial conditions and noise  are 
denoted with $\llangle\ldots \rrangle_{\tau_0}$ and $\llangle\ldots \rrangle$ respectively,
while the average over both fluctuations is denoted with the double 
brackets $\dlangle \ldots \drangle$.
Since  the transverse momentum per
rapidity is conserved for boost invariant fields, approximately boost invariant initial fluctuations in $\tau T^{\tau i}$ remain important at late times.  In \Sect{icflucts} we study initial transverse momentum 
fluctuations, while in remainder of the paper we complete our study of
thermal fluctuations during a  Bjorken expansion.

\subsection{A  Bjorken expansion with initial transverse momentum 
fluctuations}
\label{icflucts}

After the initial passage of two large nuclei in a specific event,
each rapidity interval contains a finite amount of transverse momentum, 
although the event-averaged transverse momentum per rapidity is zero.
This initial transverse momentum  is spread
over a large rapidity range by the subsequent re-scatterings in the initial
state. 
Ultimately, this dynamical process can be described by (transverse) momentum diffusion in rapidity, and can be modeled with hydrodynamics and noise -- see \Sect{sketch}.
Here we will determine how long-range transverse momentum 
fluctuations in the initial state influence the evolution of the background energy density at late times.

As a model for the initial conditions in the $x,y$ plane,
we take Gaussian statistics for 
the initial transverse momentum fluctuations
\begin{align}
    \label{gabic}
    \llangle \tau_0 g^{i}_\perp(\tau_0,\xp)  \; \tau_0 g^{j}_\perp(\tau_0, \yp) 
    \rrangle_{\tau_0} =   \chit
    \delta^{ij} \delta^{2}(\xp  - \yp), 
\end{align}
where $g^{i}_\perp(\tau, \xp) \equiv T^{\tau i}$ is approximately independent 
of 
rapidity, so that each (large) rapidity interval is approximately boost 
invariant. 
Integrating over the transverse area $\A$, the total transverse momentum  per rapidity,
\st
\frac{d p^{x}}{d\eta}  \equiv  \int_{\A} d^2x_\perp  \; \tau_0 
g^{x}_\perp(\tau_0, \xp) ,
\stp
fluctuates from event to event with a scaled variance 
of
\st
\label{chi0pp}
\chit \equiv \Big\langle \frac{1}{\A} \left(\frac{d p^{x}}{d\eta} \right)^2 \Big\rangle_{\tau_0} .
\stp

To find out how this fluctuating initial condition changes the evolution of
the system, we linearize the equations of motion of viscous hydrodynamics 
and Fourier transform with respect to the transverse coordinates
\st
\vec{g}_\perp(\tau, \kp ) \equiv \int d^2x_\perp e^{i \kp \cdot \xp }  \, \vec{g}_\perp(\tau, \xp ).
\stp
The full equations of motion are given in the next section, see \Eq{bjpert}.
Decomposing the transverse momentum fluctuation into longitudinal and transverse pieces
\st
g_\perp^i(\tau, \kp) =  g_{L}^i(\kp)  \; + \; g_{T}^{i}(\kp), 
\stp
with $\hat k^i_{\perp} g_{T}^{i} = 0$ and $g_{L}^i = \hat k^i_\perp \hat k^j_\perp g_\perp^j$,
we find that
the transverse piece obeys a two dimensional diffusion equation
\st
\partial_{\tau} (\tau g_T^i) + \gamma_{\eta} k^2_\perp (\tau g_T^i) = 0,
\stp
with initial conditions  specified by \Eq{gabic}
\begin{multline}
\label{T1T1correlatorIC}
\llangle \tau_0 g_T^i(\tau_0,\kp) \, \tau_0 g_T^j(\tau_0,-\kp') \rrangle_{\tau_0}= \\
\chit \, (\delta^{ij} - \hat k^i \hat k^j)\, (2\pi)^2 \delta^{2}(\kp - \kp').
\end{multline}
Solving the diffusion equation with a time dependent 
diffusion constant $\gamma_\eta \propto \tau^{c_s^2}$, we see that the
variance at a specified space time point due to the fluctuating initial conditions is\footnote{Here we are neglecting the longitudinal contribution, $\llangle g_{L} g_{L} \rrangle$, which decreases more rapidly than $1/\tau$ at late times. }
\begin{align}
    \llangle \tau g^i_\perp (\tau,\xp) \tau g^j_\perp(\tau,\xp) 
    \rrangle_{\tau_0}  &=  
     \delta^{ij} \frac{\chit}{12\pi\gamma_\eta \tau}  \, .
\end{align}
Thus, we see that a fluctuating initial conditions contributes
quadratically to the average stress tensor 
\begin{subequations}
\begin{align}
    \frac{\llangle \tau^2 T^{\eta\eta} \rrangle_{\tau_0}}{e + p} &= \frac{p}{e 
    +p} -  \frac{4 \gamma_\eta}{3 \tau}, \\
    \frac{\llangle T^{xx} \rrangle_{\tau_0}}{e + p} &= \frac{p}{e +p} + 
    \frac{2\gamma_\eta}{3\tau}  + \left[
\frac{\chit}{\tau^2 (e+p)^2 } \right]
    \frac{1}{12 \pi \gamma_\eta \tau}, \\
    \llangle T^{yy} \rrangle_{\tau_0}  &= \llangle T^{xx} \rrangle_{\tau_0} , \\
    \llangle T^{\tau\tau} \rrangle_{\tau_0}  &= \llangle T^{xx} 
    \rrangle_{\tau_0} + \llangle T^{yy} \rrangle_{\tau_0} + \llangle \tau^2 
    T^{\eta\eta} \rrangle_{\tau_0},
\end{align}
\end{subequations}
where $p = c_s^2 e$.

\subsection{Kinetic equations of hydrodynamic fluctuations}
\label{bjkinetic}

To derive the kinetic equations we will follow the 
strategy of \Sect{grav1}, and expand all fluctuations in Fourier modes 
conjugate to transverse coordinates and rapidity,  e.g.
\st
\delta e(\tau, \k) \equiv \int d\eta\, d^2 x_\perp \, e^{i \vec{k}_{\perp} 
\cdot \vec{x}_{\perp} + i \kappa \eta } \, \delta e(\tau, x_\perp, \eta).
\stp
The linearized equations of motion of all hydrodynamic fields around the Bjorken background read
\begin{subequations}
\label{bjpert}
\begin{align}
    0&=\left(\frac{\partial}{\partial \tau} + 
    \frac{1+c_{s}^2}{\tau}\right)\delta e
+ i\vec k_{\perp}\cdot\vec g_{\perp} + i\kappa g^{\eta} +\xi^{\tau},\\
\vec 0_{\perp}&=\left(\frac{\partial}{\partial \tau} + \frac{1}{\tau}\right) 
\vec g_{\perp}
+ c_{s}^2 i\vec k_{\perp}\delta e + \gamma_{\eta}\left(k_{\perp}^2 +  
\frac{\kappa^2}{\tau^2}\right) \vec g_{\perp}  \nonumber \\
&+\frac{1}{3}\gamma_{\eta}\vec k_{\perp}\left(\vec k_{\perp}\cdot \vec 
g_{\perp} + \kappa g^{\eta}\right) +\vec\xi_{\perp},\\
0&=\left(\frac{\partial}{\partial \tau} + \frac{3}{\tau}\right) g^{\eta}
+ \frac{c_{s}^2 i \kappa}{\tau^2}\delta e + 
\gamma_{\eta}\left(k_{\perp}^2 + \frac{\kappa^2}{\tau^2}\right) 
g^{\eta}\nonumber \\
&+ \frac{1}{3\tau^2}\gamma_{\eta}\kappa
\left(\vec k_{\perp}\cdot\vec g_{\perp} + \kappa g^{\eta}\right) + \xi^{\eta}.
\end{align}
\end{subequations}
where $(g_\perp^x, g_\perp^y, g^{\eta}) = (T^{\tau x}, T^{\tau y}, T^{\tau \eta})$.
As in \Sect{grav1} and \ref{grav2} the hydrodynamic parameters in these equations (such as $\gamma_{\eta}$) are constructed from the bare parameters, $e_0(\Lambda)$, $p_0(\Lambda)$, $\eta_0(\Lambda)$ and
evolve according to ideal hydrodynamics, $e_0(\tau) = e_0(\tau_0) (\tau_0/\tau)^{1+ c_s^2}$.
We also neglected variation in viscosity $\delta \eta/\tau \ll \delta 
p,\delta e$, which is smaller by a factor
$\epsilon=\eta_0/((e_0+p_0)c_s^2\tau)\ll 1$ for conformal fluid.
Note also that the temporal noise component $\xi^{\tau}$ is smaller than 
$\xi^{i_{\perp}}$ 
and $\tau\xi^{\eta}$ by a factor $1/(k_*\tau)\sim \epsilon^{1/2}$ and the 
former can be neglected.

Following the procedure outlined in \Sect{gravitysec} we rewrite \Eqs{bjpert} 
in a compact matrix notation. We define $\vec G=(G^{\hat x},G^{\hat y},G^{\hat 
z})\equiv(\vec g_{\perp}, \tau g^{\eta})$ and $\vec K =(K_{\hat x},K_{\hat 
y},K_{\hat z})\equiv (\vec 
k_{\perp}, \kappa/\tau)$, so that equation of motion for $\phi_a\equiv 
(c_{s}\delta e,\vec G)$ is

\begin{equation}
\label{eq:fluct_ev}
\partial_\tau\phi_a(\tau, \k)
=-i\mathcal{L}_{ab}\phi_b - \mathcal{D}_{ab}\phi_b - \xi_a - 
\mathcal{P}_{ab}\phi_b,\\
\end{equation}
\begin{align}
\mathcal{L} &=
\begin{pmatrix}
0 & c_s\vec K\\
c_s\vec K & 0
\end{pmatrix},\quad
\mathcal{D} =\gamma_\eta \begin{pmatrix}
0 & 0\\
0 & K^2\delta_{\hat \imath \hat \jmath}
+ \frac{1}{3}K_{\hat \imath}K_{\hat \jmath}
\end{pmatrix}, \\
\mathcal{P} &= \frac{1}{\tau}
\begin{pmatrix}
1+c_s^2 & & &\\
 & 1 & & \\
 & & 1 & \\
 & & & 2
\end{pmatrix},
\end{align}
with noise correlator
\begin{multline}
\langle \xi_a(\tau,\k)\xi_b(\tau',-\k) \rangle =
\frac{2T(e_0+p_0)}{\tau} \mathcal{D}_{ab}(2\pi)^3 \\
  \times \delta^3(\k-\k') \delta(\tau - \tau').
\end{multline}
Here $\delta^3(\k - \k') \equiv \delta^2(\vec{k}_\perp - \vec{k}'_{\perp}) \delta(\kappa - \kappa')$ and 
the factor of $1/\tau$ stems from the Jacobian of the coordinate
system  $\delta^4(x - x') /\sqrt{g(x)}$.

The kinetic equation for the two-point functions
\begin{equation}
\left<\phi_a(\tau,\k) 
\phi_b(\tau,-\k')\right>\equiv 
N_{ab}(\tau,\k)(2\pi)^3\delta^3(\k-\k'),
\end{equation}
is obtained similarly to \Sect{gravitysec}
\begin{equation}
\partial_\tau N(\tau,\k)
= -i[\mathcal L, N] - \{\mathcal D, N\} + \frac{2T(e_0+p_0)}{\tau}\mathcal D
-\{\mathcal P, N\}.
\end{equation}
The eigenvectors of $\mathcal{L}$ are of the same form as before, \Eq{vectors}, 
\begin{gather}
    (e_\pm)_a {=} \frac{1}{\sqrt{2}}
    \begin{pmatrix}
1\\
\pm\hat K
\end{pmatrix},\ \
(e_{T_1})_a {=} 
    \begin{pmatrix}
0\\
\vec T_1
\end{pmatrix},\ \
(e_{T_2})_a = 
    \begin{pmatrix}
0\\
\vec T_2
\end{pmatrix}.
\end{gather}
However, now the wavenumber vector $\vec{K}$
 is time dependent
\begin{subequations}
\begin{align}
    \hat{K}\equiv&  \frac{ (\vec{k}_\perp, \kappa/\tau)}{\sqrt{k_\perp^2 + (\kappa/\tau)^2 } } \,, \\ 
    \equiv &(\sin\theta_K\cos\varphi_K,\sin\theta_K\sin\varphi_K,\cos\theta_K)  \, ,
\end{align}
\end{subequations}
The azimuthal angle $\varphi_K$ is independent of time due to 
 the residual rotational symmetry of the background in $xy$\nobreakdash-plane.
Following the same arguments as in \Sect{gravitysec}, we arrive at the kinetic 
equations for diagonal components
\begin{subequations}
\label{eq:kin_bj}
\begin{align}
\label{eq:kin_N++}
\partial_\tau N_{\pm\pm}
=& -\frac{4}{3}\gamma_\eta K^2
\left[N_{\pm\pm} - \frac{T(e_0+p_0)}{\tau} \right]\nonumber \\
& \quad -\frac{1}{\tau}\left(2+c_{s}^2+\cos^2\theta_K \right)N_{\pm\pm},\\
\label{eq:kin_N11}
\partial_\tau N_{T_1T_1}
=& -2\gamma_\eta K^2 
\left[N_{T_1T_1} - \frac{T(e_0+p_0)}{\tau}\right]
-\frac{2}{\tau}N_{T_1T_1},\\
\label{eq:kin_N22}
\partial_\tau N_{T_2T_2}
=& -2\gamma_\eta K^2
\left[N_{T_2T_2} - \frac{T(e_0+p_0)}{\tau}\right] \nonumber \\
& \quad -\frac{2}{\tau}\left(1+\sin^2\theta_K\right)N_{T_2T_2}.
\end{align}
\end{subequations}
The first terms on the right hand side describe relaxation of $N_{AA}$ toward  local equilibrium $T(e_0+p_0)/\tau$, and the second terms drive $N_{AA}$ out of equilibrium through the interaction with the background flow.

We derived these equations relying on the scale separation given in
\Eq{scalesep2}.  The off-diagonal components between gapped modes (such as 
between 
the $\pm$ and $T_1$ and $T_2$ modes) are ignored because they rapidly rotate as 
discussed in \Sect{grav1}.
Note that the transverse mode $\phi_{T_1}$ is so chosen that it does not mix with the other modes.
This is possible because of the residual rotational symmetry in the $xy$-plane in the Bjorken expansion.
Therefore the kinetic equation for $N_{T_1T_1}$, \Eq{eq:kin_N11}, holds without 
the scale separation in \Eq{scalesep2} and is applicable for all wavenumbers 
$k$ from to zero to $1/\ell_{\rm mfp}$.

\subsection{Nonlinear fluctuations in the energy momentum tensor}
\label{bjstress}

\begin{figure}
\centering
\includegraphics{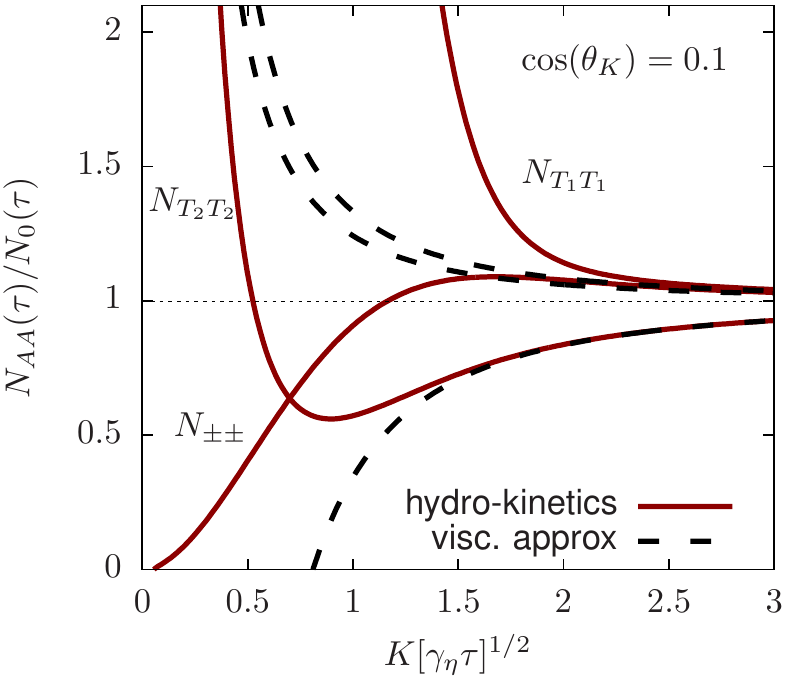}
\caption{Steady state solutions of \Eq{eq:kin_bj} for the two point 
energy-momentum correlation functions during a Bjorken 
expansion at late times, $\tau\gg \tau_0$. The correlations are plotted 
as a function of $K[\gamma_\eta \tau]^{1/2}$
for final time angle
$\cos\theta_K=0.1$.
For comparison leading order viscous solutions in $1/(\gamma_\eta K^2 \tau)$
are also shown, \Eq{eq:NAB_asmpt}. The differences of the steady state 
solutions from their asymptotic forms 
induces finite corrections 
to energy-momentum tensor, \Eq{eq:Tmunu_final}. }
\label{fig:NAA}
\end{figure}

Now let us investigate the solution of the kinetic equations close to the 
cutoff and isolate the UV divergent contribution. Solving \Eq{eq:kin_bj} in 
series of $1/(\gamma_\eta K^2\tau)$ we obtain an asymptotic solution for large 
$K/k_*$
\begin{subequations}
\label{eq:NAB_asmpt}
\begin{align}
\label{eq:N++_asmpt}
\frac{N_{\pm\pm}(\tau,\k)}{T(e_0+p_0)/\tau}&= 
1+ \frac{c_s^2 - \cos^2\theta_K}{\frac{4}{3}\gamma_{\eta} K^2\tau} 
+\ldots,\\
\label{eq:N11_asmpt}
\frac{N_{T_1T_1}(\tau,\k)}{T(e_0+p_0)/\tau}&=
1+ \frac{c_s^2}{\gamma_{\eta} K^2\tau}  +\ldots,\\
\label{eq:N22_asmpt}
\frac{N_{T_2T_2}(\tau,\k)}{T(e_0+p_0)/\tau}&=
1+ \frac{c_s^2 - \sin^2\theta_K}{\gamma_{\eta} K^2\tau} 
+ \ldots,
\end{align}
\end{subequations}
where we used $\partial_{\tau}[T(e_0+p_0)]\simeq 
-(1+2c_{s}^2)[T(e_0+p_0)]/\tau$ 
which is adequate for the desired accuracy of the present analysis.
For a given $K^2\gamma_\eta \tau = (K/k_*)^2$ and $\theta_K$ at final time 
$\tau$, we can solve \Eq{eq:kin_bj} numerically and find a steady state 
solution at late time 
$\tau \gg \tau_0$. We compare this steady state solution to the 
asymptotic form \Eq{eq:NAB_asmpt} in \Fig{fig:NAA}.

\Eq{eq:NAB_asmpt} is analogous to the ideal and first  viscous correction to 
the thermal distribution function, $f_0 + \delta f$, which are used in heavy ion
phenomenology and in determining the shear viscosity~\cite*{[see for example: 
]Teaney:2009qa}. At
large $K/k_*$ the  distribution $N_{AA}$ attains its equilibrium value,
$T(e_0 + p_0)/\tau$, up to viscous corrections of order $\tau_R/\tau$, where $\tau_{R}$ is a typical relaxation time for a mode of momentum $K$, $\tau_R \sim  1/\gamma_\eta K^2$.

The energy-momentum tensor averaged over fluctuations is given by
\begin{subequations}
\label{eq:Tmunu_tot}
\begin{align}
\label{eq:Ttt_tot}
\langle T^{\tau\tau}\rangle
&= e_0 + \frac{\langle \vec G^2\rangle}{e_0+p_0}, \\
\label{eq:Txx_tot}
\langle T^{xx}\rangle
&= p_0 + \frac{2\eta_0}{3\tau} + \frac{\langle (G^{\hat x})^2\rangle}{e_0+p_0}, \\
\label{eq:Tyy_tot}
\langle T^{yy}\rangle
&= p_0 + \frac{2\eta_0}{3\tau} + \frac{\langle (G^{\hat y})^2\rangle}{e_0+p_0},  \\
\label{eq:Tzz_tot}
\langle \tau^2 T^{\eta\eta}\rangle
&=p_0-\frac{4\eta_0}{3\tau} + \frac{\langle (G^{\hat z})^2\rangle}{e_0+p_0}.
\end{align}
\end{subequations}
Calculating $N_{ab}(\tau,\k) = \left[(e_A)_a N_{AB} (e_B)_b\right]$ from 
the kinetic theory, we determine $\langle\phi_a(\tau,\k)\phi_b(\tau,-\k)\rangle$ with 
$\phi_a = \left(c_s\delta e, \vec G\right)$ in Fourier 
space, yielding
\begin{align}
\langle\phi_a(x)\phi_b(x)\rangle
&={\int} \frac{d^2 k_{\perp}d\kappa}{(2\pi)^3} N_{ab}(\tau,\k), \nonumber \\
&= {\tau}{\int} \frac{K^2 dK d\cos \theta_K d\varphi_K}{(2\pi)^3}  \nonumber \\
  &   \qquad \times (e_A)_a N_{AB}(\tau,\k) (e_B)_{b} \label{eq:NLfluct}\, .
\end{align}
Note that the momentum integral is done in final time variables, $\vec{K}(\tau)$.
As shown below, the integration in Fourier space is divergent in the ultraviolet.
Therefore, we  regulate the integral by introducing a cutoff at $|\vec K|\sim \Lambda$.
In turn, the background quantities such as $e_0$ and $\eta_0$ must be renormalized and depend on $\Lambda$ so that the total result is independent of $\Lambda$.
The choice of $\Lambda$ is arbitrary as long as $k_{*} \ll \Lambda \ll 1/\ell_{\rm  mfp}$ so that the non-linear contribution with  $|\vec K|\sim \Lambda$ is independent of the background flow.

The integration in \Eq{eq:NLfluct} includes the soft fluctuations for which the kinetic equation may not be applicable.
However, this contribution is suppressed by phase space and
the kinetic result can be extrapolated into this regime with negligible errors.

Combining Eqs.~\eqref{eq:NAB_asmpt}, \eqref{eq:Tmunu_tot} and 
\eqref{eq:NLfluct},  the energy momentum tensor is 
obtained as 
\begin{subequations}
\begin{align}
\langle T^{\tau\tau}\rangle  &= e_0 + 3T \int_{0}^{\Lambda}\frac{K^2dK}{2\pi^2} + \Delta T^{\tau\tau}, \\
\langle T^{xx}\rangle  &= p_0 + \frac{2\eta_0}{3\tau}
+ T \int_{0}^{\Lambda}\frac{K^2dK}{2\pi^2} \\
&+\frac{17}{90}\frac{T(e_0+p_0)}{\eta_0\tau}
\int_{0}^{\Lambda}\frac{dK}{2\pi^2}  + \Delta T^{xx}, \nonumber \\
\langle T^{yy}\rangle  &= p_0 + \frac{2\eta_0}{3\tau}
+ T \int_{0}^{\Lambda}\frac{K^2dK}{2\pi^2} \\
&+\frac{17}{90}\frac{T(e_0+p_0)}{\eta_0\tau}
\int_{0}^{\Lambda}\frac{dK}{2\pi^2}  + \Delta T^{yy}, \nonumber \\
\langle \tau^2 T^{\eta\eta}\rangle  &= p_0 -\frac{4\eta_0}{3\tau}
+T\int_0^{\Lambda}\frac{K^2dK}{2\pi^2} \\
&-\frac{17}{45}\frac{T(e_0+p_0)}{\eta_0\tau}
\int_0^{\Lambda}\frac{dK}{2\pi^2} + \tau^2 \Delta T^{\eta\eta}, \nonumber 
\end{align}
\end{subequations}
where the finite contributions $\Delta T^{\tau\tau}$, $\Delta T^{xx}$, $\Delta T^{yy}$, and $\tau^2\Delta T^{\eta\eta}$ are discussed in the next section.
By comparing terms with the same explicit $\tau$ dependence, the ultraviolet 
divergences are absorbed into the renormalized hydrodynamic variables
\begin{subequations}
\label{eq:renormalization}
\begin{align}
e &= e_0(\Lambda) + \frac{T\Lambda^3}{2\pi^2}, \\
p &= p_0(\Lambda) + \frac{T\Lambda^3}{6\pi^2}, \\
\eta &= \eta_0(\Lambda) + \frac{17\Lambda}{120\pi^2}\frac{T(e_0(\Lambda)+p_0(\Lambda))}{\eta_0(\Lambda)}.
\end{align}
\end{subequations}
Note that we do not assign a cut-off dependence to the temperature. 
The coefficients of the cubic and linear renormalizations of the pressure
and shear viscosity are independent of the background expansion, and match 
the static fluid results of \Sect{grav2}.
Here $e$, $p$, and $\eta$ are physical quantities at a given temperature $T$ in an infinite volume.
Using the physical quantities, the energy-momentum tensor is given as
\begin{subequations}
\label{renTij}
\begin{align}
\langle T^{\tau\tau}(\tau)\rangle &=e + \Delta T^{\tau\tau},\\
\langle T^{xx}(\tau)\rangle &=p + \frac{2\eta}{3\tau} + \Delta T^{xx},\\
\langle T^{yy}(\tau)\rangle &=p + \frac{2\eta}{3\tau} + \Delta T^{yy},\\
\langle \tau^2T^{\eta\eta}(\tau)\rangle  &=p - \frac{4\eta}{3\tau} + \tau^2\Delta T^{\eta\eta}.
\end{align}
\end{subequations}
If the two-point functions of the fluctuations were completely determined by 
the first two terms in \Eq{eq:NAB_asmpt}, their contributions would be completely absorbed by 
the renormalization of the background flow parameters such as $p_0(\Lambda)$ and $\eta_0(\Lambda)$.
However, the kinetic equations yield residual contributions, since the full 
solution deviates from its asymptotic form for $K \sim k_*$ as seen from \Fig{fig:NAA}. The purpose of hydrodynamics with noise is to capture this contribution.

Physically, the parameters $e_0(\Lambda)$, $p_0(\Lambda)$, and $\eta_0(\Lambda)$
in fluctuating hydrodynamics 
 reflect the equilibrium properties of modes above a cutoff $\Lambda$, which have been already integrated out.
Equivalently, these parameters are determined by modes contained in a cell of size $a\sim 2\pi/\Lambda$.
For example, $p_0(\Lambda)$ is the partial pressure from equilibrated modes above the
cutoff (inside a cell), while the partial pressure from the modes below the cutoff
(larger than a cell size) is determined dynamically with fluctuating
hydrodynamics.
The second terms on the right hand sides of \Eq{eq:renormalization} are the 
contributions to each quantity from the modes below the cutoff, when all
of these long wavelength
modes are in perfect equilibrium in infinite volume.


\subsection{Out of equilibrium noise contributions to energy momentum tensor}
\label{sketch}
In this section we determine the residual contributions to the energy momentum 
tensor, $\Delta T^{\mu\nu}$, after the hydrodynamic parameters have been renormalized. We evaluate the precise numerical factors of the long-time tail terms for a 
Bjorken expansion (which is the main result of this paper), and identify 
additional contributions from  the noise at early times.
The mathematical procedure is somewhat involved, so here we outline the
calculation and present results, delegating the technical details to the 
Appendix \ref{app:finite}.

To find the full out of equilibrium correlators  we need to 
solve  \Eq{eq:kin_bj}, which can be written in the following general form
\begin{equation}\label{eq:formal}
\partial_\tau N_{AA}(\tau,\k) = f(\tau,\k) N_{AA}(\tau,\k) + g(\tau,\k),
\end{equation}
where  $f(\tau,\k)$ has contributions from both the dissipative and external 
forcing terms, and $g(\tau,\k)$ is the inhomogeneous term coming from 
the equilibrium correlation functions.
A formal solution of \Eq{eq:formal} is given by
\begin{align}
\label{eq:formalsol}
N_{AA}(\tau,\k) &= N_{AA}(\tau_0,\k) e^{\int_{\tau_0}^\tau d\tau' 
f(\tau',\k)} \nonumber\\
&+\int_{\tau_0}^\tau d\tau'' g(\tau'',\k) e^{\int_{\tau_0}^{\tau''} d\tau' 
f(\tau',\k)}.
\end{align}
The first term describes the evolution of the initial correlation density
matrix $N_{AA}(\tau_0,\k)$ to final time $\tau$. 
The second term in \Eq{eq:formalsol} is  
the contribution from thermal fluctuations.
As we will see, only the $N_{T_1T_1}$ contribution is sensitive 
to the initial conditions and the thermal fluctuations at early times.
For the $T_1T_1$ correlator we will take the initial conditions described by \Eq{T1T1correlatorIC} in \Sect{icflucts},  $\tau_0^2 N_{T_1,T_1}(\tau_0,\k) = \chit \,2\pi \delta(\kappa)$.

Substituting the formal solution for $N_{AA}$ in \Eq{eq:Tmunu_tot} and \Eq{eq:NLfluct} 
we can determine the stress tensor at time $\tau \gg \tau_0$. 
 The integral $\int d^3k$ 
in \Eq{eq:NLfluct} diverges, but after 
subtracting $\Lambda^3$ and $\Lambda$ divergences  
discussed
in the previous section, we obtain 
the finite correction to energy momentum tensor

\begin{subequations}
\begin{widetext}
\label{eq:Tmunu_final}
\begin{align}
\label{eq:Tzz_finite}
\frac{\dlangle \tau^2 T^{\eta\eta}(\tau) \drangle}{e+p}
&= \frac{p}{e + p} - \frac{4 \gamma_\eta}{3\tau} 
  + \frac{1.08318}{s\, (4\pi \gamma_{\eta}\tau)^{3/2} }, \\
\label{eq:Txxyy_finite}
\frac{\dlangle T^{xx}(\tau) \drangle}{e+p}
&=  \frac{p}{e + p } + \frac{2 \gamma_\eta}{3\tau} + 
\left[ \frac{\chit{+}\delta \chit}{\tau^2 (e+p)^2} \right]  \frac{1}{(12\pi \gamma_\eta \tau)} 
-\,  \frac{0.273836}{s \; (4\pi \gamma_\eta \tau)^{3/2} },  \\
 \dlangle T^{yy} \drangle &= \dlangle T^{xx} \drangle, \\
\label{eq:Ttt_finite}
\dlangle T^{\tau\tau}\drangle &= \dlangle T^{xx} \drangle + \dlangle T^{yy} \drangle + \dlangle \tau^2 
T^{\eta\eta} \drangle,
\end{align}
\end{widetext}
\end{subequations}
with numerically determined coefficients of the long-time tails, $1/(\gamma_\eta 
\tau)^{3/2}$. Here 
$\chit+\delta \chit$ records the initial variance
in transverse momentum in a given rapidity slice (see \Eq{gabic} and \Eq{chi0pp}) 
together with 
the thermal contribution
\st
\label{chi0ppcorrected}
\chit {+} \delta \chit = \llangle \frac{1}{\A} \left( 
\frac{dp^x}{d\eta}\right)^{\!2} \rrangle_{\!\!\tau_0} +  \left(\frac{T (e + p) 
\tau_0}{\sqrt{12\pi\gamma_\eta/\tau_0}} \right)_{\!\tau_0}  ,
\stp
where the brackets $\left(\ldots \right)_{\tau_0}$ indicate
that all contained quantities are to be evaluated at the initial time, $\tau_0$.
We will discuss the result in the
next section.
\subsection{Qualitative discussion of \Eq{eq:Tmunu_final}}
\label{qualitative}
\subsubsection{Long time tails: $1/(\gamma_\eta \tau)^{3/2}$}
\label{qual_long_time}

Examining \Eq{eq:Tmunu_final} we see two groups of terms.  
The first group is proportional
to $1/(\gamma_\eta \tau)^{3/2}$ and  is
independent of initial conditions.  By contrast, the second group is proportional to
$1/(\gamma_\eta \tau)$, and depends on the  initial transverse momentum
fluctuations through the parameter $\chit + \delta \chit$ (see 
\Sect{icflucts}).  We will  first describe the terms
proportional to the fractional power $1/(\gamma_\eta \tau)^{3/2}$,  known 
as the long-time tails.

Squared fluctuations in equilibrium are of order $\langle\delta e(\vec 
x)\delta e(\vec y)\rangle_{\rm eq}/e^2 \sim \langle v^i(\vec x) v^j(\vec 
y)\rangle_{\rm eq}\sim s^{-1}\delta(\vec x-\vec y)$.
Thus a fluctuation with wavenumber $k$ is suppressed by $\sqrt{k^3/s}$.
The suppression factor $k^3/s$ is roughly the inverse of the degrees of 
freedom inside a box of volume $ \Delta V  \sim (1/k)^3$, which must be a huge number for local thermodynamics to apply.
This is why the linear analysis of the hydrodynamic fluctuations is justified.

The energy momentum tensor in  viscous hydrodynamics is expanded  in powers of
gradients, leading to corrections in powers of $\epsilon\equiv
\eta/(e+p)\tau\ll1$.  In addition, the  fluctuations with wavenumber of
order
$|\vec K|\sim k_*$ provide a nonlinear correction to the stress tensor, which 
is  suppressed by $s \Delta V \equiv
s/k_*^3\gg 1$.  
This correction to the longitudinal pressure reflects the equipartition of energy, with $\half T$ of energy per mode, and the number of non-equilibrium modes per volume $\sim k_{*}^3$.
To summarize, the reasoning in this paragraph leads to the following parametric estimate for the longitudinal stress  
\bg
\frac{\langle\tau^2T^{\eta\eta}\rangle}{e+p}\sim 
\left[\frac{1}{4}+ \frac{\eta}{(e +p) \tau}  + \frac{1}{s (\gamma_\eta \tau)^{3/2}} 
+\cdots\right],
\nd
which is reflected  by  \Eq{eq:Tmunu_final}.

\subsubsection{ Transverse momentum diffusion in rapidity: $1/\gamma_{\eta} \tau$}
\label{diffusion_rapidity}

Additional corrections to the stress in \Eq{eq:Tmunu_final} decrease as $1/\gamma_\eta \tau$, in contrast to the long time tails.
As described in \Sect{icflucts}, long range (in rapidity) initial transverse momentum fluctuations 
correct the mean transverse pressures, $T^{xx}$ and $T^{yy}$,
by a term proportional to  $\chit/\gamma_\eta \tau$ (see \Eq{eq:Txxyy_finite}).
Hydrodynamic noise in the initial state adjusts this correction by adding to
the long range fluctuations of transverse momentum (see \Eqs{eq:Txxyy_finite} 
and \eq{chi0ppcorrected}).  The 
goal of this section is to explain this process qualitatively, and to quantitatively
explain the adjustment, $\chit \rightarrow \chit + \delta\chit$.

Formally, the $N_{T_1T_1}$ correlation function is 
sensitive to the noise at the initial time $\tau_0$, which arises
from a 
restricted region  of $\vec{K}$-space integration, $k_{\perp}\sim k_*$ 
and $ \kappa/\tau \sim k_{*}(\tau_0) (\tau_0/\tau) \sim k_*
(\tau_0/\tau)^{1/3}\ll k_*$. In this region the longitudinal momentum $\kappa/\tau_0$
reflects the dissipative scale $k_{*}(\tau_0)$ at the initial time $\tau_0$, while
the transverse momenta reflect the dissipative scale at final time $\tau$.

The dynamics in this phase space region is the following. During the initial
moments, thermal fluctuations lead to a local fluctuation of  transverse momentum in a given rapidity slice for each cell in the
transverse plane  
\st
\llangle (\tau_0 \Delta g^x_\perp)^2 \rrangle \sim  \left(\frac{T\, (e +p) 
\,\tau_0 }{\Delta\eta \, (\Delta x_\perp)^2 } \right)_{\tau_0}.
\stp
Here (as before) the brackets $(\ldots)_{\tau_0}$ indicate that all contained quantities should be evaluated at $\tau_0$.
During an initial time of order $\tau_0$, the momentum per rapidity diffuses to a finite 
longitudinal width~\cite{Gavin:2006xd} (see below)
\st
\Delta \eta   \rightarrow   \sigma_{\eta}(\tau_0) \equiv \sqrt{6\gamma_\eta(\tau_0)/\tau_0} .
\stp
The process is diffusive because the transverse momentum per rapidity is 
conserved.
The rapidity width is finite because the longitudinal expansion 
shuts off the diffusion process.
$\sigma_\eta(\tau_0)$ is broader than the rapidity width of subsequent interest, which is of order $\sigma_\eta(\tau)$. Thus, 
after an initial transient, the transverse momentum 
per rapidity may be considered approximately constant in time and rapidity, though localized in this  transverse plane
\st
\llangle (\tau \Delta g^x_\perp)^2  \rrangle \sim   \left(\frac{T\, (e +p) 
\,\tau_0 }{\sqrt{\gamma_\eta/\tau_0} } \right)_{\tau_0} 
\frac{1}{(\Delta x_\perp)^2 }. 
\stp
At much later times these transverse momentum fluctuations diffuse transversely (as described in \Sect{icflucts}) leading to a correction of order
\st
\frac{ \llangle T^{xx} \rrangle }{e + p } \sim 
\frac{1}{\tau^2 (e + p)^2 } \left( \frac{T (e +p) \tau_0 }{\sqrt{\gamma_\eta/\tau_0} } \right)_{\tau_0} \frac{1}{\gamma_\eta \tau} ,
\stp
which qualitatively reproduces the correction in \Eq{eq:Txxyy_finite}.

Now we will briefly sketch this reasoning with equations.
At the early time moments $\tau \sim \tau_0$, the wave vector is predominantly longitudinal 
$\vec{K} \simeq (\vec{0}_\perp, \kappa/\tau)$ and the 
transverse momentum correlator
\st
\llangle g^i_{\perp}(\tau,\k) g^{j}_{\perp}(\tau,-\k')\rrangle 
\equiv  N^{ij}(\k,\tau) (2\pi)^3 \delta^3(\k - \k') ,
\stp
can be reconstructed from $N_{T_1T_1}$ and $N_{T_2T_2}$
\st
N^{ij}(\tau,\k) = \sum_{A \in T_1, T_2} e_A^{i} e_A^{j} N_{AA}(\tau,\k),
\stp
since $\vec{T}_1$ and $\vec{T}_2$ form a basis for the transverse plane.
In this limit, the equations of motion for $N_{T_1T_1}$ and $N_{T_2T_2}$ (see \Eq{eq:kin_N11} and \Eq{eq:kin_N22}) are the same,
and $N^{ij}$ satisfies a one dimensional  diffusion equation with a source at
early times
\st
\label{longdiffuse}
\left( \partial_{\tau} +  2\gamma_{\eta} \left(\frac{\kappa}{\tau}\right)^2\right) (\tau^2 N^{ij} ) = 2\gamma_\eta \left(\frac{\kappa}{\tau}\right)^2 \, T(e+p)\tau \delta^{ij} \, .
\stp
The lhs of \Eq{longdiffuse} represents the diffusion 
of transverse momentum
in rapidity,
 while the rhs represents the thermal transverse momentum fluctuations at
the earliest moments, which act as a source.
The source for the fluctuations, $2 T \eta \, (\kappa/\tau)^2$, 
is a rapidly decreasing function of time, and is dominant for times of order $\tau_0$.

The Green function 
propagating data from $\tau'$ to $\tau$ 
for the lhs of \Eq{longdiffuse} 
is  
\begin{align}
    G^{ij}(\tau\eta\xp|\tau'\eta'\xp') {=}
\frac{
e^{- (\eta - \eta')^2/(12 \gamma_\eta(\tau')/\tau')  }  
}{\sqrt{12\pi \gamma_\eta(\tau')/\tau'}} 
\delta^{ij} \delta^2(\xp - \xp'),
\end{align}
for $\tau \gg \tau'$.
Thus, a fluctuation localized in rapidity at time $\tau_0$ will diffuse to a 
finite rapidity width  of $\sigma_{\eta}(\tau_0)=\sqrt{6 
\gamma_\eta(\tau_0)/\tau_0}$  at late times\footnote{In 
\Refs{Gavin:2006xd,Gavin:2016hmv} the authors consider an initial 
distribution 
which is Gaussian in rapidity of width $\sigma_0$. During 
the expansion the width
is broadened by the diffusion process
\st
\label{gavinresult}
\sigma^2_0 \rightarrow \sigma_0^2 + 6  \frac{\gamma_\eta(\tau_0) }{\tau_0}\,.
\stp
These authors considered constant $\eta/(e + p)$ 
and found a factor of $4$ rather than $6$ in \Eq{gavinresult}.
}~\cite{Gavin:2006xd,Gavin:2016hmv}.
This 
is a small rapidity width in absolute units (since $\gamma_{\eta}(\tau_0)/\tau_0 \ll 1$ when hydrodynamics is a good approximation), but much broader than the rapidity
width of interest at the final time, $\gamma_\eta(\tau_0)/\tau_0 \gg \gamma_\eta(\tau)/\tau$.

Returning to \Eq{longdiffuse}, we solve the equation,
and determine the transverse momentum correlation function 
(in
the same rapidity slice)  
at an intermediate time $\tau'$ which 
is large compared to $\tau_0$ but much  
much less than the final time $\tau$, $\tau_0 \ll \tau' \ll \tau$
\begin{multline}
\tau'^2 \llangle g^{i}_\perp(\tau',\eta, \xp) g^{j}_\perp(\tau',\eta, \yp) 
\rrangle 
= \\ 
\int  \frac{d\kappa d^2k_\perp}{(2\pi)^3 } e^{i\vec{k}_\perp \cdot (\xp - \yp)  }
     \; \tau'^2 N^{ij}(\tau',\kappa) \, .
\end{multline}
Implementing these steps we find
\begin{multline}
\tau'^2 \llangle g^{i}_\perp(\tau',\eta,\xp) g^{j}_\perp(\tau',\eta, \yp) 
\rrangle = \\ 
\left(\frac{T(e +p))\tau_0}{\sqrt{ 12\pi \gamma_\eta/\tau_0 
}}\right)_{\!\!\tau_0}\!\! \delta^{ij} \delta^2(\xp - \yp).
\end{multline}
This has the same form as the initial conditions described in \Sect{icflucts},
and fluctuations at the earliest moments simply increase the variance of
long range transverse momentum fluctuations by a constant amount
\st
\delta \chit =  \left(\frac{T (e + p) \tau_0}{\sqrt{12\pi\gamma_\eta/\tau_0}} 
\right)_{\!\tau_0},
\stp
reproducing \Eq{chi0ppcorrected}.
In a sense, this constant shift simply finalizes the thermalization process
described at the start of \Sect{icflucts}.  The correction $\delta \chit$ scales as
$\tau_0^{-1/3}$ and is therefore small compared to the first
term in \Eq{chi0ppcorrected} if $\tau_0$ is large compared to a typical thermalization time.


\section{Results and Discussion}
\label{results}
In this paper we determined a set of kinetic equations which describe
the evolution of hydrodynamic fluctuations during a Bjorken expansion.
We used these equations to
find the first fractional power correction 
to the longitudinal pressure, $\propto 1/(\tau T)^{3/2}$, at late times.
The evolution equations can be extended to much more general flows, 
and ultimately coupled to existing hydrodynamic codes.

The kinetic equations for hydrodynamic fluctuations 
are a WKB (or rotating wave) type approximation of the full stochastic hydrodynamic evolution
equations. This approximation
is justified because the relevant hydrodynamic modes have
wavenumbers of order
\st
k_{*}  \sim \sqrt{ \frac{e  + p}{\eta \tau } } \, ,
\stp
which is large compared to the inverse expansion rate, $1/\tau$. 
For example, the kinetic equation for the sound mode 
with wavenumber $\vec{K} = (\vec{k}_\perp , \kappa/\tau)$
interacting
with the Bjorken background takes the form of 
a relaxation type equation 
\begin{multline}
    \label{soundpp}
    \partial_\tau N_{++}(\tau,\k)
= -\frac{4}{3}\gamma_\eta K^2
\left[N_{++} - \frac{T(e_0+p_0)}{\tau} \right]  \\
-\frac{1}{\tau}\left(2+c_{s}^2+ \cos\theta_K \right)N_{++} \, .
\end{multline}
$N_{++}(\tau,\k)$ are  short wavelength (symmetrized) two point
functions of conserved stress tensor components, $\phi_+ \equiv (c_s \delta e + 
\hat{K} 
\cdot \vec{G})/\sqrt{2}$ in an evolving Bjorken hydrodynamic background (see 
\Sect{bjorken} and \Eq{eq:kin_bj} for the remaining modes).
At high wavenumbers $K \gg k_{*}$, the distribution function $N_{++}$ reaches
its equilibrium form $T (e_0+ p_0)/\tau$, up to first viscous
corrections  which may be found by solving \Eq{soundpp} order by order at large
$K/k_*$ (see \Eq{eq:N++_asmpt}).   This asymptotic 
form is responsible for the renormalization of the pressure and shear viscosity.
For wavenumbers of order $k_{*}$ the hydrodynamic
fluctuations are not in equilibrium at all, but reach a non-equilibrium steady
state at late times.  A graph of this non-equilibrium steady state is given in 
\Fig{fig:NAA}.  

The deviation of hydrodynamic fluctuations from equilibrium has
consequences for the evolution of the system. Indeed, the 
longitudinal pressure $\tau^2 T^{\eta\eta}$ receives a correction
from the unequilibrated modes 
\begin{multline}
    \label{tzzrepeat2}
\frac{\llangle \tau^2T^{\eta\eta}\rrangle }{e + p} =   \Big[ \frac{p}{e+p} -  
    \frac{4}{3} \frac{\gamma_\eta}{\tau}  
+     \frac{1.08318}{s \, (4\pi \gamma_\eta \tau)^{3/2} }  \\ +   \frac{(\lambda_1 - \eta\tau_\pi)}{e+p} \frac{8}{9\tau^2} \Big] \, ,
\end{multline}
where we have repeated \Eq{eq:Tzz_finite} for convenience.  The 
correction to the pressure ${\sim}T/(\gamma_\eta \tau)^{3/2}$ is of order 
${\sim}T k_{*}^3$, reflecting
the number of modes of order $k_{*}$ and the energy 
per mode, $\half T$.
In contrast to all previous analyses of long-time tails~\cite{Kovtun:2003vj,Kovtun:2011np},
the hydrodynamic fluctuations in the expanding case are not close to 
equilibrium, and a loop expansion around equilibrium is not an appropriate approximation scheme.

Formally, the noise correction  is lower order than the correction due 
to second order hydrodynamics, which is proportional to a particular combination of second order parameters, $\lambda_1 - \eta \tau_\pi$. To quantify the importance 
of thermal fluctuations in practice,
we take representative numbers 
for the entropy from
the lattice~\cite{Borsanyi:2013bia,Bazavov:2014pvz}, estimates 
for the second order hydrodynamic coefficients
based on weakly and strongly coupled plasmas~\cite{york:2008rr,Bhattacharyya:2008jc,Baier:2007ix},
and an estimate for $\tau T$ at $\tau\sim 3.5\,{\rm fm}$ based on hydrodynamic
simulations\footnote{We take an estimate for the (approximately constant)
average entropy in the transverse plane from a recent LHC simulation for PbPb
collisions at $\sqrt{s}=2.76\,{\rm TeV}/{\rm nucleon}$, $\llangle \tau_{o}
s(\tau_o) \rrangle \simeq 4.0\,{\rm GeV}^2$~\cite{Mazeliauskas:2015vea}.  We 
take a time of $\tau \sim 3.5\,{\rm fm}$ (which is the time at which the 
elliptic flow develops~\cite*{[see
for example: ]Teaney:2009qa}), where $T\simeq 250\,{\rm MeV}$. } 
\begin{subequations}
\begin{align}
    \frac{T^3}{s} \simeq&  \frac{1}{13.5}  \,, \\
    \frac{(\lambda_1 - \eta\tau_\pi)}{e+p} \simeq& -0.8 \left(\frac{\eta}{e+p} \right)^2  \,, \\
     \tau T  \simeq 4.5 \, .
\end{align}
\end{subequations}
Then, for $\eta/s \simeq 1/4\pi$, \Eq{tzzrepeat2} evaluates to
\begin{multline}
    \label{oneby4pi}
\frac{ \llangle \tau^2 T^{\eta\eta} \rrangle }{e+ p} = \frac{1}{4} \Big[
    1.-0.092\,\left(\frac{4.5}{\tau T}\right)+0.034\,\left(\frac{4.5}{\tau T}\right)^{3/2} \\
-0.00085 \left(\frac{4.5}{\tau T}\right)^{2} \Big],
\end{multline}
while for $\eta/s = 2/4\pi$, we find
\begin{multline}
    \label{twoby4pi}
\frac{ \llangle \tau^2 T^{\eta\eta} \rrangle }{e+ p} = \frac{1}{4} \Big[
    1.-0.185 \left(\frac{4.5}{\tau T}\right) + 0.013\left(\frac{4.5}{\tau T}\right)^{3/2}  \\
   -0.0034\left(\frac{4.5}{\tau T}\right)^{2} \Big].
\end{multline}
For the smaller shear viscosity, \Eq{oneby4pi}, 
the 
nonlinear noise contribution completely dominates over
the second order hydro contribution.    For the larger shear
viscosity, \Eq{twoby4pi}, the noise remains three times larger
than  second order hydro, but this contribution is 
 only a $\sim 10\%$ of the first order viscous term.

The evolution of the average energy density of the system obeys
\st
\frac{d \dlangle T^{\tau\tau} \drangle}{d\tau} = -\frac{ \dlangle T^{\tau\tau}\drangle  +  \dlangle \tau^2T^{\eta\eta} \drangle}{\tau},\ \ \
\stp
where the double brackets notate an average over (long range in rapidity)
initial conditions and thermal noise\footnote{The
    longitudinal pressure in \Eq{tzzrepeat2} is independent of fluctuations in
    the initial conditions at late times. Thus, only 
    the average over the noise is relevant in this case,
$\dlangle \tau^2 T^{\eta\eta} \drangle = \llangle \tau^2 T^{\eta\eta} \rrangle$ }. To close the system of equations, the relationship between average energy density $\dlangle T^{\tau\tau} \drangle$ and the average rest frame energy  density $e(\tau)$ must be specified, and this relation is given in \Eq{eq:Tmunu_final}.  
$T^{\tau\tau}$, $T^{xx}$,  and $T^{yy}$  are sensitive
to hydrodynamic noise at the earliest moments in addition
to the long-time tails.
In these cases thermal noise in the initial state
adds to the long-range  rapidity correlation functions of transverse momentum, which are already present without noise. 
This result is encapsulated by \Eq{chi0ppcorrected} and is discussed in
\Sect{icflucts} and \Sect{diffusion_rapidity}.

Although the analysis of hydrodynamic fluctuations in this paper was limited to conformal neutral fluids and a Bjorken expansion, the techniques developed here can be  applied to much 
more general flows, such as the Hubble  expansion.
In addition, it will be phenomenologically important to extend this work to non-conformal systems with net baryon number. Near the QCD critical point the noise will continue to grow without
bound, leading  to a critical renormalization of the  bulk
viscosity. In an expanding system these fluctuations will not be fully
equilibrated. We believe the formalism set up in this paper provides the first
steps towards quantitatively analyzing this rich dynamical regime.

\begin{acknowledgments}
We thank Li Yan for contributions during the initial stages of this work.
We would also like to thank the Institute for Nuclear Theory at the University 
of Washington, where this work was initiated during the  
workshop on ``Correlations and Fluctuations in p+A and A+A Collisions".
Y.A. is supported by JSPS Postdoctoral Fellowships for Research Abroad.
A.M. and D.T. work was supported in part by the U.S. Department of Energy under 
Contract No. DE\nobreakdash-FG\nobreakdash-88ER40388.
\end{acknowledgments}

\appendix

\section{Computation of finite residual contributions}
\label{app:finite}
In this appendix we provide the details of the computation sketched in \Sect{sketch} 
for the residual out of equilibrium noise contribution  to the energy momentum tensor 
for a Bjorken background.
Let us scale the correlation density matrix  by the equilibrium 
value:
\begin{align}
R_{AA}(\tau,\k)\equiv \frac{N_{AA}(\tau,\k)}{T(e_0+p_0)/\tau}.
\end{align}
The kinetic equations of motion \Eq{eq:kin_bj} written for relative density 
matrix $R_{AA}$ are
\begin{align}
\partial_{\tau}R_{\pm\pm}
&= -\frac{4}{3}\gamma_{\eta}K^2
\left(R_{\pm\pm} -1\right)
+\frac{c_s^2-\cos^2\theta_K}{\tau}R_{\pm\pm},\\
\partial_{\tau}R_{T_1T_1}
&= -2\gamma_{\eta}K^2 \left(R_{T_1T_1} - 1\right)
+\frac{2c_s^2}{\tau}R_{T_1T_1},\\
\partial_{\tau}R_{T_2T_2}
&= -2\gamma_{\eta}K^2 \left(R_{T_2T_2} - 1\right)
+\frac{2(c_s^2-\sin^2\theta_K)}{\tau} R_{T_2T_2}.
\end{align}
Using dimensionless variables $t\equiv \tau'/\tau$ and $\vec r\equiv \vec K/k_*$ with $\vec K=(\vec k_{\perp},\kappa/\tau)$ and 
$k_*=1/(\gamma_\eta\tau)^{1/2}$ defined at $\tau$, the Green functions for 
the homogeneous parts are
\begin{align}
\label{eq:green_sound}
G_{\pm\pm}(\tau',\tau;\k)
&=\frac{1}{t^{c_s^2}}\frac{1}{\sqrt{A(t,\theta_K)}}\exp\left[-\frac{4}{3}r^2 B(t,\theta_K)\right], \\
\label{eq:green_t1}
G_{T_1T_1}(\tau',\tau;\k)
&=\frac{1}{t^{2c_s^2}}\exp\left[
-2r^2 B(t,\theta_K)\right],\\
\label{eq:green_t2}
G_{T_2T_2}(\tau',\tau;\k)
&= t^{2-2c_s^2}A(t,\theta_K)
\exp\left[-2r^2B(t,\theta_K)\right],
\end{align}
where
\begin{align}
A(t,\theta_K)&\equiv\sin^2\theta_K + \frac{\cos^2\theta_K}{t^2}, \\
B(t,\theta_K)&\equiv
\frac{\sin^2\theta_K}{1+c_s^2}
\left(1-t^{1+c_s^2}\right)
+\frac{\cos^2\theta_K}{1-c_s^2}
\left(\frac{1}{t^{1-c_s^2}} -1\right).
\end{align}
With these Green functions, $R_{AA}$ due to thermal fluctuations (in contrast 
to initial fluctuations discussed in \Sect{icflucts}) is given by
\begin{align}
\label{RppGreen}
R_{++}(\tau,\k)&=
 \int_{\tau_0}^{\tau}d\tau' \frac{4}{3}\gamma_{\eta}(\tau')
\left(k_{\perp}^2+\frac{\kappa^2}{\tau'^2}\right)G_{++}(\tau',\tau;\k),
\end{align}
and similarly for the other modes (change 4/3 to 2 for the 
transverse modes). 
Since the asymptotic solution of $R_{AA}$ for large $K$ is known, we define the 
remainder of $R_{AA}$ as:
\begin{align}
\label{eq:rest_sound}
R^{(\rm r)}_{\pm\pm}(\tau,\k) &\equiv R_{\pm\pm}(\tau,\k) - \left(1 + \frac{c_s^2 - \cos^2\theta_K}{\frac{4}{3}\gamma_{\eta} K^2\tau}\right), \\
\label{eq:rest_t1}
R^{(\rm r)}_{T_1T_1}(\tau,\k) &\equiv R_{T_1T_1}(\tau,\k) - \left(1+ \frac{c_s^2}{\gamma_{\eta} K^2\tau}\right), \\
\label{eq:rest_t2}
R^{(\rm r)}_{T_2T_2}(\tau,\k) &\equiv R_{T_2T_2}(\tau,\k) - \left(1+ \frac{c_s^2 - \sin^2\theta_K}{\gamma_{\eta} K^2\tau}\right).
\end{align}

Using $R_{AA}^{(\rm r)}$ the residual contribution to the energy-momentum 
tensor is calculated from \Eq{eq:Ttt_tot} and \Eq{eq:NLfluct} as 
\begin{align}
\label{eq:DeltaTxx}
\Delta T^{xx} &= T
\int\frac{d^3K}{(2\pi)^3}\left[
\begin{aligned}
&\frac{R^{(\rm r)}_{++}+R^{(\rm r)}_{--}}{2} \sin^2\theta_K\cos^2\varphi_K \\
&+ R^{(\rm r)}_{T_1T_1}\sin^2\varphi_K \\
&+ R^{(\rm r)}_{T_2T_2}\cos^2\theta_K\cos^2\varphi_K 
\end{aligned}
\right],
\end{align}
\begin{align}
\label{eq:DeltaTyy}
\Delta T^{yy} &= T
\int\frac{d^3K}{(2\pi)^3}\left[
\begin{aligned}
&\frac{R^{(\rm r)}_{++}+R^{(\rm r)}_{--}}{2} \sin^2\theta_K\sin^2\varphi_K \\
&+ R^{(\rm r)}_{T_1T_1}\cos^2\varphi_K \\
&+ R^{(\rm r)}_{T_2T_2}\cos^2\theta_K\sin^2\varphi_K 
\end{aligned}
\right],
\end{align}
\begin{align}
\label{eq:DeltaTzz}
\tau^2 \Delta T^{\eta\eta} &= T
\int\frac{d^3K}{(2\pi)^3}\left[
\begin{aligned}
&\frac{R^{(\rm r)}_{++}+R^{(\rm r)}_{--}}{2} \cos^2\theta_K \\
&+ R^{(\rm r)}_{T_2T_2}\sin^2\theta_K
\end{aligned}
\right],
\end{align}
\begin{align}
\label{eq:DeltaTtt}
\Delta T^{\tau\tau} &= \Delta T^{xx} + \Delta T^{yy} + \tau^2\Delta T^{\eta\eta}.
\end{align}
Substituting the subtracted solution $R^{(\rm r)}_{AA}$ into 
\eqref{eq:DeltaTxx}-\eqref{eq:DeltaTtt} and performing $r$ integration with a 
Gaussian cutoff  $\exp[-r^2k_*^2/\Lambda^2]$, we get
\begin{align}
&\frac{\left[\tau^2\Delta T^{\eta\eta}(\tau)\right]}{T(\tau) k_*^3}
= \frac{3\sqrt{\pi}}{8}\int_{-1}^1\frac{d(\cos\theta_K)}{4\pi^2}\int_{\tau_0/\tau\to 0}^1 dt \nonumber \\
& \ \ \ \times\left(
\frac{\frac{4}{3}\cos^2\theta_K\sqrt{A(t,\theta_K)}}
{\left[\frac{4}{3}B(t,\theta_K) + k_*^2/\Lambda^2\right]^{5/2}}
+ \frac{2t^{2-c_{s0}^2}\sin^2\theta_K  A(t,\theta_K)^2}
{\left[2B(t,\theta_K) + k_*^2/\Lambda^2\right]^{5/2}}
\right) \nonumber \\
& \ \ \ -\left[\mathcal O(\Lambda^3) + \mathcal O(\Lambda)\right],
\end{align}
\begin{align}
&\frac{\left[\Delta T^{xx}(\tau)+\Delta T^{yy}(\tau)\right]}{T(\tau) k_*^3}
= \frac{3\sqrt{\pi}}{8}\int_{-1}^1\frac{d(\cos\theta_K)}{4\pi^2}\int_{\tau_0/\tau}^1 dt \nonumber \\
& \ \ \ \ \ \ \times\left(
\frac{\frac{4}{3}\sin^2\theta_K\sqrt{A(t,\theta_K)}}
{\left[\frac{4}{3}B(t,\theta_K) + k_*^2/\Lambda^2\right]^{5/2}}\right. \nonumber \\
& \ \ \ \ \ \ \ \ \ \ \ \left. 
+ \ \frac{2t^{2-c_s^2}\cos^2\theta_K A(t,\theta_K)^2
+ 2t^{-c_s^2}A(t,\theta_K)}
{\left[2B(t,\theta_K) + k_*^2/\Lambda^2\right]^{5/2}}
\right) \nonumber \\
& \ \ \ \ \ -\left[\mathcal O(\Lambda^3) + \mathcal O(\Lambda)\right] 
\end{align}
The ultraviolet divergent terms $\mathcal O(\Lambda^3,\Lambda)$ are from the asymptotic form of $R_{AA}$ at large $K$ in \eqref{eq:rest_sound}-\eqref{eq:rest_t2}.
Near $t=1$,  $B(t,\theta_K)\simeq 1-t$ and the cutoff $\Lambda$ regulates the 
divergence in time integral.
To isolate the divergences, we perform the partial integration twice and pick up cubic and linear divergences from the surface terms at $t=1$.
The resultant divergences are precisely canceled by $\mathcal O(\Lambda^3,\Lambda)$ terms.

After subtracting the ultraviolet divergences at $t=1$ and doing $\cos 
\theta_K$ integral analytically, the remaining time integration has to be done 
numerically.  $R^{(\rm r)}_{T_1T_1}$ mode contribution to 
$T^{xx}$ and $T^{yy}$ is divergent in the limit $\tau \gg \tau_0$.
Since the analytic behavior of the integrand around $t\sim 0$ is known, we can 
explicitly subtract the part sensitive to early times from the integrand to 
extract remaining 
finite pieces for $R^{(\rm r)}_{T_1T_1}$ mode. Numerical integration results 
necessary to find 
finite 
stress tensor corrections in \Eqs{eq:DeltaTxx}-\eq{eq:DeltaTzz} are summarized 
in 
Table \ref{tab1}. 
Summing contributions 
from the different modes to the longitudinal and transverse components of energy 
momentum tensor gives the numerical coefficients $1.08318$ and $-0.273836$ as seen 
in \Eq{eq:Tmunu_final}.

\begin{table}[t]
\def\arraystretch{1.5}
$
\begin{array}{|c|c|c|}
\hline
R^{(\rm r)}_{AA} & (4\pi)^{-3/2}\int d^3r R^{(\rm r)}_{AA} & (4\pi)^{-3/2}\int 
d^3r \cos^2 
\theta_K R^{(\rm r)}_{AA}\\
\hline
R^{(\rm r)}_{\pm\pm} & -0.439511 & 0.021281 \\ 
R^{(\rm r)}_{T_1T_1} & -\frac{\pi}{3\sqrt{6}}\approx-0.427517 & -0.467513 \\ 
R^{(\rm r)}_{T_2T_2} & 1.402539 & 0.340636\\
\hline
\end{array}
$
\caption{\label{tab1} Numerical values of finite pieces of regularized $R^{(\rm 
r)}_{AA}$ integrals for energy momentum tensor corrections. For the special 
case of $\int d^3r R^{(\rm r)}_{T_1T_1}$ the remaining one dimensional time 
integral can be done 
analytically.} 
\end{table}

\bibliography{master}

\begin{thebibliography}{34}%
\makeatletter
\providecommand \@ifxundefined [1]{%
 \@ifx{#1\undefined}
}%
\providecommand \@ifnum [1]{%
 \ifnum #1\expandafter \@firstoftwo
 \else \expandafter \@secondoftwo
 \fi
}%
\providecommand \@ifx [1]{%
 \ifx #1\expandafter \@firstoftwo
 \else \expandafter \@secondoftwo
 \fi
}%
\providecommand \natexlab [1]{#1}%
\providecommand \enquote  [1]{``#1''}%
\providecommand \bibnamefont  [1]{#1}%
\providecommand \bibfnamefont [1]{#1}%
\providecommand \citenamefont [1]{#1}%
\providecommand \href@noop [0]{\@secondoftwo}%
\providecommand \href [0]{\begingroup \@sanitize@url \@href}%
\providecommand \@href[1]{\@@startlink{#1}\@@href}%
\providecommand \@@href[1]{\endgroup#1\@@endlink}%
\providecommand \@sanitize@url [0]{\catcode `\\12\catcode `\$12\catcode
  `\&12\catcode `\#12\catcode `\^12\catcode `\_12\catcode `\%12\relax}%
\providecommand \@@startlink[1]{}%
\providecommand \@@endlink[0]{}%
\providecommand \url  [0]{\begingroup\@sanitize@url \@url }%
\providecommand \@url [1]{\endgroup\@href {#1}{\urlprefix }}%
\providecommand \urlprefix  [0]{URL }%
\providecommand \Eprint [0]{\href }%
\providecommand \doibase [0]{http://dx.doi.org/}%
\providecommand \selectlanguage [0]{\@gobble}%
\providecommand \bibinfo  [0]{\@secondoftwo}%
\providecommand \bibfield  [0]{\@secondoftwo}%
\providecommand \translation [1]{[#1]}%
\providecommand \BibitemOpen [0]{}%
\providecommand \bibitemStop [0]{}%
\providecommand \bibitemNoStop [0]{.\EOS\space}%
\providecommand \EOS [0]{\spacefactor3000\relax}%
\providecommand \BibitemShut  [1]{\csname bibitem#1\endcsname}%
\let\auto@bib@innerbib\@empty
\bibitem [{\citenamefont {Bjorken}(1983)}]{Bjorken:1982qr}%
  \BibitemOpen
  \bibfield  {author} {\bibinfo {author} {\bibfnamefont {J.~D.}\ \bibnamefont
  {Bjorken}},\ }\bibfield  {title} {\enquote {\bibinfo {title} {{Highly
  Relativistic Nucleus-Nucleus Collisions: The Central Rapidity Region}},}\
  }\href {\doibase 10.1103/PhysRevD.27.140} {\bibfield  {journal} {\bibinfo
  {journal} {Phys. Rev.}\ }\textbf {\bibinfo {volume} {D27}},\ \bibinfo {pages}
  {140--151} (\bibinfo {year} {1983})}\BibitemShut {NoStop}%
\bibitem [{\citenamefont {Heinz}\ and\ \citenamefont
  {Snellings}(2013)}]{Heinz:2013th}%
  \BibitemOpen
  \bibfield  {author} {\bibinfo {author} {\bibfnamefont {Ulrich}\ \bibnamefont
  {Heinz}}\ and\ \bibinfo {author} {\bibfnamefont {Raimond}\ \bibnamefont
  {Snellings}},\ }\bibfield  {title} {\enquote {\bibinfo {title} {{Collective
  flow and viscosity in relativistic heavy-ion collisions}},}\ }\href {\doibase
  10.1146/annurev-nucl-102212-170540} {\bibfield  {journal} {\bibinfo
  {journal} {Ann. Rev. Nucl. Part. Sci.}\ }\textbf {\bibinfo {volume} {63}},\
  \bibinfo {pages} {123--151} (\bibinfo {year} {2013})},\ \Eprint
  {http://arxiv.org/abs/1301.2826} {arXiv:1301.2826 [nucl-th]} \BibitemShut
  {NoStop}%
\bibitem [{\citenamefont {Luzum}\ and\ \citenamefont
  {Petersen}(2014)}]{Luzum:2013yya}%
  \BibitemOpen
  \bibfield  {author} {\bibinfo {author} {\bibfnamefont {Matthew}\ \bibnamefont
  {Luzum}}\ and\ \bibinfo {author} {\bibfnamefont {Hannah}\ \bibnamefont
  {Petersen}},\ }\bibfield  {title} {\enquote {\bibinfo {title} {{Initial State
  Fluctuations and Final State Correlations in Relativistic Heavy-Ion
  Collisions}},}\ }\href {\doibase 10.1088/0954-3899/41/6/063102} {\bibfield
  {journal} {\bibinfo  {journal} {J. Phys.}\ }\textbf {\bibinfo {volume}
  {G41}},\ \bibinfo {pages} {063102} (\bibinfo {year} {2014})},\ \Eprint
  {http://arxiv.org/abs/1312.5503} {arXiv:1312.5503 [nucl-th]} \BibitemShut
  {NoStop}%
\bibitem [{\citenamefont {Baier}\ \emph {et~al.}(2008)\citenamefont {Baier},
  \citenamefont {Romatschke}, \citenamefont {Son}, \citenamefont {Starinets},\
  and\ \citenamefont {Stephanov}}]{Baier:2007ix}%
  \BibitemOpen
  \bibfield  {author} {\bibinfo {author} {\bibfnamefont {Rudolf}\ \bibnamefont
  {Baier}}, \bibinfo {author} {\bibfnamefont {Paul}\ \bibnamefont
  {Romatschke}}, \bibinfo {author} {\bibfnamefont {Dam~Thanh}\ \bibnamefont
  {Son}}, \bibinfo {author} {\bibfnamefont {Andrei~O.}\ \bibnamefont
  {Starinets}}, \ and\ \bibinfo {author} {\bibfnamefont {Mikhail~A.}\
  \bibnamefont {Stephanov}},\ }\bibfield  {title} {\enquote {\bibinfo {title}
  {{Relativistic viscous hydrodynamics, conformal invariance, and
  holography}},}\ }\href {\doibase 10.1088/1126-6708/2008/04/100} {\bibfield
  {journal} {\bibinfo  {journal} {JHEP}\ }\textbf {\bibinfo {volume} {04}},\
  \bibinfo {pages} {100} (\bibinfo {year} {2008})},\ \Eprint
  {http://arxiv.org/abs/0712.2451} {arXiv:0712.2451 [hep-th]} \BibitemShut
  {NoStop}%
\bibitem [{\citenamefont {Gavin}\ and\ \citenamefont
  {Abdel-Aziz}(2006)}]{Gavin:2006xd}%
  \BibitemOpen
  \bibfield  {author} {\bibinfo {author} {\bibfnamefont {Sean}\ \bibnamefont
  {Gavin}}\ and\ \bibinfo {author} {\bibfnamefont {Mohamed}\ \bibnamefont
  {Abdel-Aziz}},\ }\bibfield  {title} {\enquote {\bibinfo {title} {{Measuring
  Shear Viscosity Using Transverse Momentum Correlations in Relativistic
  Nuclear Collisions}},}\ }\href {\doibase 10.1103/PhysRevLett.97.162302}
  {\bibfield  {journal} {\bibinfo  {journal} {Phys. Rev. Lett.}\ }\textbf
  {\bibinfo {volume} {97}},\ \bibinfo {pages} {162302} (\bibinfo {year}
  {2006})},\ \Eprint {http://arxiv.org/abs/nucl-th/0606061}
  {arXiv:nucl-th/0606061 [nucl-th]} \BibitemShut {NoStop}%
\bibitem [{\citenamefont {Kapusta}\ \emph {et~al.}(2012)\citenamefont
  {Kapusta}, \citenamefont {Muller},\ and\ \citenamefont
  {Stephanov}}]{Kapusta:2011gt}%
  \BibitemOpen
  \bibfield  {author} {\bibinfo {author} {\bibfnamefont {J.~I.}\ \bibnamefont
  {Kapusta}}, \bibinfo {author} {\bibfnamefont {B.}~\bibnamefont {Muller}}, \
  and\ \bibinfo {author} {\bibfnamefont {M.}~\bibnamefont {Stephanov}},\
  }\bibfield  {title} {\enquote {\bibinfo {title} {{Relativistic Theory of
  Hydrodynamic Fluctuations with Applications to Heavy Ion Collisions}},}\
  }\href {\doibase 10.1103/PhysRevC.85.054906} {\bibfield  {journal} {\bibinfo
  {journal} {Phys. Rev.}\ }\textbf {\bibinfo {volume} {C85}},\ \bibinfo {pages}
  {054906} (\bibinfo {year} {2012})},\ \Eprint {http://arxiv.org/abs/1112.6405}
  {arXiv:1112.6405 [nucl-th]} \BibitemShut {NoStop}%
\bibitem [{\citenamefont {Yan}\ and\ \citenamefont
  {Gr{\"o}nqvist}(2016)}]{Yan:2015lfa}%
  \BibitemOpen
  \bibfield  {author} {\bibinfo {author} {\bibfnamefont {Li}~\bibnamefont
  {Yan}}\ and\ \bibinfo {author} {\bibfnamefont {Hanna}\ \bibnamefont
  {Gr{\"o}nqvist}},\ }\bibfield  {title} {\enquote {\bibinfo {title}
  {{Hydrodynamical noise and Gubser flow}},}\ }\href {\doibase
  10.1007/JHEP03(2016)121} {\bibfield  {journal} {\bibinfo  {journal} {JHEP}\
  }\textbf {\bibinfo {volume} {03}},\ \bibinfo {pages} {121} (\bibinfo {year}
  {2016})},\ \Eprint {http://arxiv.org/abs/1511.07198} {arXiv:1511.07198
  [nucl-th]} \BibitemShut {NoStop}%
\bibitem [{\citenamefont {Young}\ \emph {et~al.}(2015)\citenamefont {Young},
  \citenamefont {Kapusta}, \citenamefont {Gale}, \citenamefont {Jeon},\ and\
  \citenamefont {Schenke}}]{Young:2014pka}%
  \BibitemOpen
  \bibfield  {author} {\bibinfo {author} {\bibfnamefont {C.}~\bibnamefont
  {Young}}, \bibinfo {author} {\bibfnamefont {J.~I.}\ \bibnamefont {Kapusta}},
  \bibinfo {author} {\bibfnamefont {C.}~\bibnamefont {Gale}}, \bibinfo {author}
  {\bibfnamefont {S.}~\bibnamefont {Jeon}}, \ and\ \bibinfo {author}
  {\bibfnamefont {B.}~\bibnamefont {Schenke}},\ }\bibfield  {title} {\enquote
  {\bibinfo {title} {{Thermally Fluctuating Second-Order Viscous Hydrodynamics
  and Heavy-Ion Collisions}},}\ }\href {\doibase 10.1103/PhysRevC.91.044901}
  {\bibfield  {journal} {\bibinfo  {journal} {Phys. Rev.}\ }\textbf {\bibinfo
  {volume} {C91}},\ \bibinfo {pages} {044901} (\bibinfo {year} {2015})},\
  \Eprint {http://arxiv.org/abs/1407.1077} {arXiv:1407.1077 [nucl-th]}
  \BibitemShut {NoStop}%
\bibitem [{\citenamefont {Murase}\ and\ \citenamefont
  {Hirano}(2016)}]{Murase:2016rhl}%
  \BibitemOpen
  \bibfield  {author} {\bibinfo {author} {\bibfnamefont {Koichi}\ \bibnamefont
  {Murase}}\ and\ \bibinfo {author} {\bibfnamefont {Tetsufumi}\ \bibnamefont
  {Hirano}},\ }\bibfield  {title} {\enquote {\bibinfo {title} {{Hydrodynamic
  fluctuations and dissipation in an integrated dynamical model}},}\ }in\ \href
  {http://inspirehep.net/record/1414798/files/arXiv:1601.02260.pdf} {\emph
  {\bibinfo {booktitle} {{Proceedings, 33rd International Symposium on Lattice
  Field Theory (Lattice 2015)}}}}\ (\bibinfo {year} {2016})\ \Eprint
  {http://arxiv.org/abs/1601.02260} {arXiv:1601.02260 [nucl-th]} \BibitemShut
  {NoStop}%
\bibitem [{\citenamefont {Nagai}\ \emph {et~al.}(2016)\citenamefont {Nagai},
  \citenamefont {Kurita}, \citenamefont {Murase},\ and\ \citenamefont
  {Hirano}}]{Nagai:2016wyx}%
  \BibitemOpen
  \bibfield  {author} {\bibinfo {author} {\bibfnamefont {Kenichi}\ \bibnamefont
  {Nagai}}, \bibinfo {author} {\bibfnamefont {Ryuichi}\ \bibnamefont {Kurita}},
  \bibinfo {author} {\bibfnamefont {Koichi}\ \bibnamefont {Murase}}, \ and\
  \bibinfo {author} {\bibfnamefont {Tetsufumi}\ \bibnamefont {Hirano}},\
  }\bibfield  {title} {\enquote {\bibinfo {title} {{Causal hydrodynamic
  fluctuation in Bjorken expansion}},}\ \ }(\bibinfo {year} {2016})\ \Eprint
  {http://arxiv.org/abs/1602.00794} {arXiv:1602.00794 [nucl-th]} \BibitemShut
  {NoStop}%
\bibitem [{\citenamefont {Bell}\ \emph {et~al.}(2007)\citenamefont {Bell},
  \citenamefont {Garcia},\ and\ \citenamefont {Williams}}]{bell2007numerical}%
  \BibitemOpen
  \bibfield  {author} {\bibinfo {author} {\bibfnamefont {John~B.}\ \bibnamefont
  {Bell}}, \bibinfo {author} {\bibfnamefont {Alejandro~L.}\ \bibnamefont
  {Garcia}}, \ and\ \bibinfo {author} {\bibfnamefont {Sarah~A.}\ \bibnamefont
  {Williams}},\ }\bibfield  {title} {\enquote {\bibinfo {title} {Numerical
  methods for the stochastic {Landau-Lifshitz Navier-Stokes} equations},}\
  }\href {\doibase 10.1103/PhysRevE.76.016708} {\bibfield  {journal} {\bibinfo
  {journal} {Phys. Rev. E}\ }\textbf {\bibinfo {volume} {76}},\ \bibinfo
  {pages} {016708} (\bibinfo {year} {2007})}\BibitemShut {NoStop}%
\bibitem [{\citenamefont {Donev}\ \emph {et~al.}(2011)\citenamefont {Donev},
  \citenamefont {Bell}, \citenamefont {de~la Fuente},\ and\ \citenamefont
  {Garcia}}]{donev2011diffusive}%
  \BibitemOpen
  \bibfield  {author} {\bibinfo {author} {\bibfnamefont {Aleksandar}\
  \bibnamefont {Donev}}, \bibinfo {author} {\bibfnamefont {John~B.}\
  \bibnamefont {Bell}}, \bibinfo {author} {\bibfnamefont {Anton}\ \bibnamefont
  {de~la Fuente}}, \ and\ \bibinfo {author} {\bibfnamefont {Alejandro~L.}\
  \bibnamefont {Garcia}},\ }\bibfield  {title} {\enquote {\bibinfo {title}
  {Diffusive transport by thermal velocity fluctuations},}\ }\href {\doibase
  10.1103/PhysRevLett.106.204501} {\bibfield  {journal} {\bibinfo  {journal}
  {Phys. Rev. Lett.}\ }\textbf {\bibinfo {volume} {106}},\ \bibinfo {pages}
  {204501} (\bibinfo {year} {2011})}\BibitemShut {NoStop}%
\bibitem [{\citenamefont {Usabiaga}\ \emph {et~al.}(2012)\citenamefont
  {Usabiaga}, \citenamefont {Bell}, \citenamefont {Delgado-Buscalioni},
  \citenamefont {Donev}, \citenamefont {Fai}, \citenamefont {Griffith},\ and\
  \citenamefont {Peskin}}]{balboa2012staggered}%
  \BibitemOpen
  \bibfield  {author} {\bibinfo {author} {\bibfnamefont {Florencio~Balboa}\
  \bibnamefont {Usabiaga}}, \bibinfo {author} {\bibfnamefont {John~B.}\
  \bibnamefont {Bell}}, \bibinfo {author} {\bibfnamefont {Rafael}\ \bibnamefont
  {Delgado-Buscalioni}}, \bibinfo {author} {\bibfnamefont {Aleksandar}\
  \bibnamefont {Donev}}, \bibinfo {author} {\bibfnamefont {Thomas~G.}\
  \bibnamefont {Fai}}, \bibinfo {author} {\bibfnamefont {Boyce~E.}\
  \bibnamefont {Griffith}}, \ and\ \bibinfo {author} {\bibfnamefont
  {Charles~S.}\ \bibnamefont {Peskin}},\ }\bibfield  {title} {\enquote
  {\bibinfo {title} {Staggered schemes for fluctuating hydrodynamics},}\ }\href
  {\doibase 10.1137/120864520} {\bibfield  {journal} {\bibinfo  {journal}
  {Multiscale Modeling \& Simulation}\ }\textbf {\bibinfo {volume} {10}},\
  \bibinfo {pages} {1369--1408} (\bibinfo {year} {2012})}\BibitemShut {NoStop}%
\bibitem [{\citenamefont {Alder}\ and\ \citenamefont
  {Wainwright}(1967)}]{velocity_auto}%
  \BibitemOpen
  \bibfield  {author} {\bibinfo {author} {\bibfnamefont {B.~J.}\ \bibnamefont
  {Alder}}\ and\ \bibinfo {author} {\bibfnamefont {T.~E.}\ \bibnamefont
  {Wainwright}},\ }\bibfield  {title} {\enquote {\bibinfo {title} {Velocity
  autocorrelations for hard spheres},}\ }\href {\doibase
  10.1103/PhysRevLett.18.988} {\bibfield  {journal} {\bibinfo  {journal} {Phys.
  Rev. Lett.}\ }\textbf {\bibinfo {volume} {18}},\ \bibinfo {pages} {988--990}
  (\bibinfo {year} {1967})}\BibitemShut {NoStop}%
\bibitem [{\citenamefont {Alder}\ and\ \citenamefont
  {Wainwright}(1970)}]{wainwright2}%
  \BibitemOpen
  \bibfield  {author} {\bibinfo {author} {\bibfnamefont {B.~J.}\ \bibnamefont
  {Alder}}\ and\ \bibinfo {author} {\bibfnamefont {T.~E.}\ \bibnamefont
  {Wainwright}},\ }\bibfield  {title} {\enquote {\bibinfo {title} {Decay of the
  velocity autocorrelation function},}\ }\href {\doibase 10.1103/PhysRevA.1.18}
  {\bibfield  {journal} {\bibinfo  {journal} {Phys. Rev. A}\ }\textbf {\bibinfo
  {volume} {1}},\ \bibinfo {pages} {18--21} (\bibinfo {year}
  {1970})}\BibitemShut {NoStop}%
\bibitem [{\citenamefont {Zwanzig}\ and\ \citenamefont
  {Bixon}(1970)}]{bixon_zwanzig}%
  \BibitemOpen
  \bibfield  {author} {\bibinfo {author} {\bibfnamefont {Robert}\ \bibnamefont
  {Zwanzig}}\ and\ \bibinfo {author} {\bibfnamefont {Mordechai}\ \bibnamefont
  {Bixon}},\ }\bibfield  {title} {\enquote {\bibinfo {title} {Hydrodynamic
  theory of the velocity correlation function},}\ }\href {\doibase
  10.1103/PhysRevA.2.2005} {\bibfield  {journal} {\bibinfo  {journal} {Phys.
  Rev. A}\ }\textbf {\bibinfo {volume} {2}},\ \bibinfo {pages} {2005--2012}
  (\bibinfo {year} {1970})}\BibitemShut {NoStop}%
\bibitem [{\citenamefont {Stephanov}\ \emph {et~al.}(1998)\citenamefont
  {Stephanov}, \citenamefont {Rajagopal},\ and\ \citenamefont
  {Shuryak}}]{Stephanov:1998dy}%
  \BibitemOpen
  \bibfield  {author} {\bibinfo {author} {\bibfnamefont {Misha~A.}\
  \bibnamefont {Stephanov}}, \bibinfo {author} {\bibfnamefont {K.}~\bibnamefont
  {Rajagopal}}, \ and\ \bibinfo {author} {\bibfnamefont {Edward~V.}\
  \bibnamefont {Shuryak}},\ }\bibfield  {title} {\enquote {\bibinfo {title}
  {{Signatures of the tricritical point in QCD}},}\ }\href {\doibase
  10.1103/PhysRevLett.81.4816} {\bibfield  {journal} {\bibinfo  {journal}
  {Phys. Rev. Lett.}\ }\textbf {\bibinfo {volume} {81}},\ \bibinfo {pages}
  {4816--4819} (\bibinfo {year} {1998})},\ \Eprint
  {http://arxiv.org/abs/hep-ph/9806219} {arXiv:hep-ph/9806219 [hep-ph]}
  \BibitemShut {NoStop}%
\bibitem [{\citenamefont {Loizides}(2016)}]{Loizides:2016tew}%
  \BibitemOpen
  \bibfield  {author} {\bibinfo {author} {\bibfnamefont {Constantin}\
  \bibnamefont {Loizides}},\ }\bibfield  {title} {\enquote {\bibinfo {title}
  {{Experimental overview on small collision systems at the LHC}},}\ }in\ \href
  {https://inspirehep.net/record/1424892/files/arXiv:1602.09138.pdf} {\emph
  {\bibinfo {booktitle} {{25th International Conference on Ultra-Relativistic
  Nucleus-Nucleus Collisions (Quark Matter 2015) Kobe, Japan, September
  27-October 3, 2015}}}}\ (\bibinfo {year} {2016})\ \Eprint
  {http://arxiv.org/abs/1602.09138} {arXiv:1602.09138 [nucl-ex]} \BibitemShut
  {NoStop}%
\bibitem [{\citenamefont {Teaney}(2009)}]{Teaney:2009qa}%
  \BibitemOpen
  \bibfield  {author} {\bibinfo {author} {\bibfnamefont {Derek~A.}\
  \bibnamefont {Teaney}},\ }\bibfield  {title} {\enquote {\bibinfo {title}
  {{Viscous Hydrodynamics and the Quark Gluon Plasma}},}\ }\href@noop {} {\
  (\bibinfo {year} {2009})},\ \Eprint {http://arxiv.org/abs/0905.2433}
  {arXiv:0905.2433 [nucl-th]} \BibitemShut {NoStop}%
\bibitem [{\citenamefont {Son}\ and\ \citenamefont
  {Starinets}(2007)}]{Son:2007vk}%
  \BibitemOpen
  \bibfield  {author} {\bibinfo {author} {\bibfnamefont {Dam~T.}\ \bibnamefont
  {Son}}\ and\ \bibinfo {author} {\bibfnamefont {Andrei~O.}\ \bibnamefont
  {Starinets}},\ }\bibfield  {title} {\enquote {\bibinfo {title} {{Viscosity,
  Black Holes, and Quantum Field Theory}},}\ }\href {\doibase
  10.1146/annurev.nucl.57.090506.123120} {\bibfield  {journal} {\bibinfo
  {journal} {Ann. Rev. Nucl. Part. Sci.}\ }\textbf {\bibinfo {volume} {57}},\
  \bibinfo {pages} {95--118} (\bibinfo {year} {2007})},\ \Eprint
  {http://arxiv.org/abs/0704.0240} {arXiv:0704.0240 [hep-th]} \BibitemShut
  {NoStop}%
\bibitem [{\citenamefont {Kovtun}\ \emph {et~al.}(2011)\citenamefont {Kovtun},
  \citenamefont {Moore},\ and\ \citenamefont {Romatschke}}]{Kovtun:2011np}%
  \BibitemOpen
  \bibfield  {author} {\bibinfo {author} {\bibfnamefont {Pavel}\ \bibnamefont
  {Kovtun}}, \bibinfo {author} {\bibfnamefont {Guy~D.}\ \bibnamefont {Moore}},
  \ and\ \bibinfo {author} {\bibfnamefont {Paul}\ \bibnamefont {Romatschke}},\
  }\bibfield  {title} {\enquote {\bibinfo {title} {{The stickiness of sound: An
  absolute lower limit on viscosity and the breakdown of second order
  relativistic hydrodynamics}},}\ }\href {\doibase 10.1103/PhysRevD.84.025006}
  {\bibfield  {journal} {\bibinfo  {journal} {Phys. Rev.}\ }\textbf {\bibinfo
  {volume} {D84}},\ \bibinfo {pages} {025006} (\bibinfo {year} {2011})},\
  \Eprint {http://arxiv.org/abs/1104.1586} {arXiv:1104.1586 [hep-ph]}
  \BibitemShut {NoStop}%
\bibitem [{\citenamefont {Hong}\ and\ \citenamefont
  {Teaney}(2010)}]{Hong:2010at}%
  \BibitemOpen
  \bibfield  {author} {\bibinfo {author} {\bibfnamefont {Juhee}\ \bibnamefont
  {Hong}}\ and\ \bibinfo {author} {\bibfnamefont {Derek}\ \bibnamefont
  {Teaney}},\ }\bibfield  {title} {\enquote {\bibinfo {title} {{Spectral
  densities for hot QCD plasmas in a leading log approximation}},}\ }\href
  {\doibase 10.1103/PhysRevC.82.044908} {\bibfield  {journal} {\bibinfo
  {journal} {Phys. Rev.}\ }\textbf {\bibinfo {volume} {C82}},\ \bibinfo {pages}
  {044908} (\bibinfo {year} {2010})},\ \Eprint {http://arxiv.org/abs/1003.0699}
  {arXiv:1003.0699 [nucl-th]} \BibitemShut {NoStop}%
\bibitem [{\citenamefont {Landau}\ and\ \citenamefont
  {Lifshitz}(1980)}]{LandauStatPart1}%
  \BibitemOpen
  \bibfield  {author} {\bibinfo {author} {\bibfnamefont {L.D.}\ \bibnamefont
  {Landau}}\ and\ \bibinfo {author} {\bibfnamefont {E.M.}\ \bibnamefont
  {Lifshitz}},\ }\href@noop {} {\emph {\bibinfo {title} {Statistical
  Physics}}},\ \bibinfo {series} {Course of theoretical physics}, Vol.~\bibinfo
  {volume} {5}\ (\bibinfo  {publisher} {Pergamon Press},\ \bibinfo {year}
  {1980})\BibitemShut {NoStop}%
\bibitem [{\citenamefont {Lifshitz}\ and\ \citenamefont
  {Pitaevskii}(1980)}]{LandauStatPart2}%
  \BibitemOpen
  \bibfield  {author} {\bibinfo {author} {\bibfnamefont {E.M.}\ \bibnamefont
  {Lifshitz}}\ and\ \bibinfo {author} {\bibfnamefont {L.P.}\ \bibnamefont
  {Pitaevskii}},\ }\href@noop {} {\emph {\bibinfo {title} {Statistical
  Physics}}},\ \bibinfo {series} {Course of theoretical physics}, Vol.~\bibinfo
  {volume} {9}\ (\bibinfo  {publisher} {Pergamon Press},\ \bibinfo {year}
  {1980})\BibitemShut {NoStop}%
\bibitem [{\citenamefont {Kovtun}\ and\ \citenamefont
  {Yaffe}(2003)}]{Kovtun:2003vj}%
  \BibitemOpen
  \bibfield  {author} {\bibinfo {author} {\bibfnamefont {Pavel}\ \bibnamefont
  {Kovtun}}\ and\ \bibinfo {author} {\bibfnamefont {Laurence~G.}\ \bibnamefont
  {Yaffe}},\ }\bibfield  {title} {\enquote {\bibinfo {title} {{Hydrodynamic
  fluctuations, long time tails, and supersymmetry}},}\ }\href {\doibase
  10.1103/PhysRevD.68.025007} {\bibfield  {journal} {\bibinfo  {journal} {Phys.
  Rev.}\ }\textbf {\bibinfo {volume} {D68}},\ \bibinfo {pages} {025007}
  (\bibinfo {year} {2003})},\ \Eprint {http://arxiv.org/abs/hep-th/0303010}
  {arXiv:hep-th/0303010 [hep-th]} \BibitemShut {NoStop}%
\bibitem [{\citenamefont {Caron-Huot}\ and\ \citenamefont
  {Saremi}(2010)}]{CaronHuot:2009iq}%
  \BibitemOpen
  \bibfield  {author} {\bibinfo {author} {\bibfnamefont {Simon}\ \bibnamefont
  {Caron-Huot}}\ and\ \bibinfo {author} {\bibfnamefont {Omid}\ \bibnamefont
  {Saremi}},\ }\bibfield  {title} {\enquote {\bibinfo {title} {{Hydrodynamic
  Long-Time tails From Anti de Sitter Space}},}\ }\href {\doibase
  10.1007/JHEP11(2010)013} {\bibfield  {journal} {\bibinfo  {journal} {JHEP}\
  }\textbf {\bibinfo {volume} {11}},\ \bibinfo {pages} {013} (\bibinfo {year}
  {2010})},\ \Eprint {http://arxiv.org/abs/0909.4525} {arXiv:0909.4525
  [hep-th]} \BibitemShut {NoStop}%
\bibitem [{\citenamefont {Danielewicz}\ and\ \citenamefont
  {Gyulassy}(1985)}]{Danielewicz:1984ww}%
  \BibitemOpen
  \bibfield  {author} {\bibinfo {author} {\bibfnamefont {P.}~\bibnamefont
  {Danielewicz}}\ and\ \bibinfo {author} {\bibfnamefont {M.}~\bibnamefont
  {Gyulassy}},\ }\bibfield  {title} {\enquote {\bibinfo {title} {{Dissipative
  Phenomena in Quark Gluon Plasmas}},}\ }\href {\doibase
  10.1103/PhysRevD.31.53} {\bibfield  {journal} {\bibinfo  {journal} {Phys.
  Rev.}\ }\textbf {\bibinfo {volume} {D31}},\ \bibinfo {pages} {53--62}
  (\bibinfo {year} {1985})}\BibitemShut {NoStop}%
\bibitem [{\citenamefont {Bhattacharyya}\ \emph {et~al.}(2008)\citenamefont
  {Bhattacharyya}, \citenamefont {Hubeny}, \citenamefont {Minwalla},\ and\
  \citenamefont {Rangamani}}]{Bhattacharyya:2008jc}%
  \BibitemOpen
  \bibfield  {author} {\bibinfo {author} {\bibfnamefont {Sayantani}\
  \bibnamefont {Bhattacharyya}}, \bibinfo {author} {\bibfnamefont {Veronika~E}\
  \bibnamefont {Hubeny}}, \bibinfo {author} {\bibfnamefont {Shiraz}\
  \bibnamefont {Minwalla}}, \ and\ \bibinfo {author} {\bibfnamefont {Mukund}\
  \bibnamefont {Rangamani}},\ }\bibfield  {title} {\enquote {\bibinfo {title}
  {{Nonlinear Fluid Dynamics from Gravity}},}\ }\href {\doibase
  10.1088/1126-6708/2008/02/045} {\bibfield  {journal} {\bibinfo  {journal}
  {JHEP}\ }\textbf {\bibinfo {volume} {02}},\ \bibinfo {pages} {045} (\bibinfo
  {year} {2008})},\ \Eprint {http://arxiv.org/abs/0712.2456} {arXiv:0712.2456
  [hep-th]} \BibitemShut {NoStop}%
\bibitem [{\citenamefont {Jeon}\ and\ \citenamefont
  {Yaffe}(1996)}]{Jeon:1995zm}%
  \BibitemOpen
  \bibfield  {author} {\bibinfo {author} {\bibfnamefont {Sangyong}\
  \bibnamefont {Jeon}}\ and\ \bibinfo {author} {\bibfnamefont {Laurence~G.}\
  \bibnamefont {Yaffe}},\ }\bibfield  {title} {\enquote {\bibinfo {title}
  {{From quantum field theory to hydrodynamics: Transport coefficients and
  effective kinetic theory}},}\ }\href {\doibase 10.1103/PhysRevD.53.5799}
  {\bibfield  {journal} {\bibinfo  {journal} {Phys. Rev.}\ }\textbf {\bibinfo
  {volume} {D53}},\ \bibinfo {pages} {5799--5809} (\bibinfo {year} {1996})},\
  \Eprint {http://arxiv.org/abs/hep-ph/9512263} {arXiv:hep-ph/9512263 [hep-ph]}
  \BibitemShut {NoStop}%
\bibitem [{\citenamefont {Gavin}\ \emph {et~al.}(2016)\citenamefont {Gavin},
  \citenamefont {Moschelli},\ and\ \citenamefont {Zin}}]{Gavin:2016hmv}%
  \BibitemOpen
  \bibfield  {author} {\bibinfo {author} {\bibfnamefont {Sean}\ \bibnamefont
  {Gavin}}, \bibinfo {author} {\bibfnamefont {George}\ \bibnamefont
  {Moschelli}}, \ and\ \bibinfo {author} {\bibfnamefont {Christopher}\
  \bibnamefont {Zin}},\ }\bibfield  {title} {\enquote {\bibinfo {title}
  {{Rapidity Correlation Structure in Nuclear Collisions}},}\ }\href@noop {} {\
   (\bibinfo {year} {2016})},\ \Eprint {http://arxiv.org/abs/1606.02692}
  {arXiv:1606.02692 [nucl-th]} \BibitemShut {NoStop}%
\bibitem [{\citenamefont {Borsanyi}\ \emph {et~al.}(2014)\citenamefont
  {Borsanyi}, \citenamefont {Fodor}, \citenamefont {Hoelbling}, \citenamefont
  {Katz}, \citenamefont {Krieg},\ and\ \citenamefont
  {Szabo}}]{Borsanyi:2013bia}%
  \BibitemOpen
  \bibfield  {author} {\bibinfo {author} {\bibfnamefont {Szabocls}\
  \bibnamefont {Borsanyi}}, \bibinfo {author} {\bibfnamefont {Zoltan}\
  \bibnamefont {Fodor}}, \bibinfo {author} {\bibfnamefont {Christian}\
  \bibnamefont {Hoelbling}}, \bibinfo {author} {\bibfnamefont {Sandor~D.}\
  \bibnamefont {Katz}}, \bibinfo {author} {\bibfnamefont {Stefan}\ \bibnamefont
  {Krieg}}, \ and\ \bibinfo {author} {\bibfnamefont {Kalman~K.}\ \bibnamefont
  {Szabo}},\ }\bibfield  {title} {\enquote {\bibinfo {title} {{Full result for
  the QCD equation of state with 2+1 flavors}},}\ }\href {\doibase
  10.1016/j.physletb.2014.01.007} {\bibfield  {journal} {\bibinfo  {journal}
  {Phys. Lett.}\ }\textbf {\bibinfo {volume} {B730}},\ \bibinfo {pages}
  {99--104} (\bibinfo {year} {2014})},\ \Eprint
  {http://arxiv.org/abs/1309.5258} {arXiv:1309.5258 [hep-lat]} \BibitemShut
  {NoStop}%
\bibitem [{\citenamefont {Bazavov}\ \emph {et~al.}(2014)\citenamefont {Bazavov}
  \emph {et~al.}}]{Bazavov:2014pvz}%
  \BibitemOpen
  \bibfield  {author} {\bibinfo {author} {\bibfnamefont {A.}~\bibnamefont
  {Bazavov}} \emph {et~al.} (\bibinfo {collaboration} {HotQCD}),\ }\bibfield
  {title} {\enquote {\bibinfo {title} {{Equation of state in ( 2+1 )-flavor
  QCD}},}\ }\href {\doibase 10.1103/PhysRevD.90.094503} {\bibfield  {journal}
  {\bibinfo  {journal} {Phys. Rev.}\ }\textbf {\bibinfo {volume} {D90}},\
  \bibinfo {pages} {094503} (\bibinfo {year} {2014})},\ \Eprint
  {http://arxiv.org/abs/1407.6387} {arXiv:1407.6387 [hep-lat]} \BibitemShut
  {NoStop}%
\bibitem [{\citenamefont {York}\ and\ \citenamefont
  {Moore}(2009)}]{york:2008rr}%
  \BibitemOpen
  \bibfield  {author} {\bibinfo {author} {\bibfnamefont {Mark~Abraao}\
  \bibnamefont {York}}\ and\ \bibinfo {author} {\bibfnamefont {Guy~D.}\
  \bibnamefont {Moore}},\ }\bibfield  {title} {\enquote {\bibinfo {title}
  {{Second order hydrodynamic coefficients from kinetic theory}},}\ }\href
  {\doibase 10.1103/PhysRevD.79.054011} {\bibfield  {journal} {\bibinfo
  {journal} {Phys. Rev.}\ }\textbf {\bibinfo {volume} {D79}},\ \bibinfo {pages}
  {054011} (\bibinfo {year} {2009})},\ \Eprint {http://arxiv.org/abs/0811.0729}
  {arXiv:0811.0729 [hep-ph]} \BibitemShut {NoStop}%
\bibitem [{\citenamefont {Mazeliauskas}\ and\ \citenamefont
  {Teaney}(2015)}]{Mazeliauskas:2015vea}%
  \BibitemOpen
  \bibfield  {author} {\bibinfo {author} {\bibfnamefont {Aleksas}\ \bibnamefont
  {Mazeliauskas}}\ and\ \bibinfo {author} {\bibfnamefont {Derek}\ \bibnamefont
  {Teaney}},\ }\bibfield  {title} {\enquote {\bibinfo {title} {{Subleading
  harmonic flows in hydrodynamic simulations of heavy ion collisions}},}\
  }\href {\doibase 10.1103/PhysRevC.91.044902} {\bibfield  {journal} {\bibinfo
  {journal} {Phys. Rev.}\ }\textbf {\bibinfo {volume} {C91}},\ \bibinfo {pages}
  {044902} (\bibinfo {year} {2015})},\ \Eprint
  {http://arxiv.org/abs/1501.03138} {arXiv:1501.03138 [nucl-th]} \BibitemShut
  {NoStop}%
\end{thebibliography}%

\end{document}